\documentclass[aps, twocolumn, superscriptaddress, longbibliography]{revtex4-1}

\usepackage{graphicx}

\usepackage{amsmath}
\usepackage{amssymb}
\usepackage{graphicx}\usepackage{afterpage}
\usepackage{amsfonts}
\usepackage{amssymb}
\usepackage{graphicx}
\usepackage{braket}
\usepackage{textcomp}

\newcommand{\Ea}{\ensuremath{{\cal E}_1}}
\newcommand{\Eb}{\ensuremath{{\cal E}_2}}
\newcommand{\Ec}{\ensuremath{{\cal E}_3}}

\newcommand{\Er}{\ensuremath{{\cal E}_{\rm R}}}
\newcommand{\Eo}{\ensuremath{{\cal E}_{1,2,3}}}

\begin{document}
\title{Coherent dynamics and mapping of excitons\\ in single-layer MoSe$_2$ and WSe$_2$ at the homogeneous limit}

\author{Caroline~Boule}
\affiliation{Univ. Grenoble Alpes, CNRS, Grenoble INP, Institut
N\'{e}el, 38000 Grenoble, France}

\author{Diana~Vaclavkova}
\affiliation{Laboratoire National des Champs Magn\'{e}tiques
Intenses, CNRS-UGA-UPS-INSA-EMFL, 25 Av. des Martyrs, 38042
Grenoble, France}

\author{Miroslav~Bartos}
\affiliation{Laboratoire National des Champs Magn\'{e}tiques
Intenses, CNRS-UGA-UPS-INSA-EMFL, 25 Av. des Martyrs, 38042
Grenoble, France} \affiliation{Central European Institute of
Technology, Brno University of Technology, Purkynova 656/123, 61200
Brno, Czech Republic}

\author{Karol~Nogajewski}
\affiliation{Faculty of Physics, University of Warsaw, ul. Pasteura
5, 02-093 Warszawa, Poland}

\author{Lukas~Zdra\v{z}il}
\affiliation{Laboratoire National des Champs Magn\'{e}tiques
Intenses, CNRS-UGA-UPS-INSA-EMFL, 25 Av. des Martyrs, 38042
Grenoble, France} \affiliation{Regional Centre of Advanced
Technologies and Materials, Department of Physical Chemistry,
Faculty of Science, Palack\'{y} University Olomouc,
\v{S}lechtitel\.{u} 27, 78371 Olomouc, Czech Republic}

\author{Takashi~Taniguchi}
\affiliation{National Institute for Materials Science, Tsukuba,
Ibaraki, 305-0044 Japan}

\author{Kenji~Watanabe}
\affiliation{National Institute for Materials Science, Tsukuba,
Ibaraki, 305-0044 Japan}

\author{Marek~Potemski}
\affiliation{Laboratoire National des Champs Magn\'{e}tiques
Intenses, CNRS-UGA-UPS-INSA-EMFL, 25 Av. des Martyrs, 38042
Grenoble, France} \affiliation{Faculty of Physics, University of
Warsaw, ul. Pasteura 5, 02-093 Warszawa, Poland}

\author{Jacek~Kasprzak}
\email{jacek.kasprzak@neel.cnrs.fr} \affiliation{Univ. Grenoble
Alpes, CNRS, Grenoble INP, Institut N\'{e}el, 38000 Grenoble,
France}


\begin{abstract}

We perform coherent nonlinear spectroscopy of excitons in
single-layers of MoSe$_2$ and WSe$_2$ encapsulated between thin
films of hexagonal boron nitride. Employing four-wave mixing
microscopy we identify virtually disorder free areas, generating
exciton optical response at the homogeneous limit. Focussing on such
areas, we measure exciton homogeneous broadening as a function of
environmental factors, namely temperature and exciton density.
Exploiting FWM imaging, we find that at such locations, nonlinear
absorption of the exciton excited states and their coherent
couplings can be observed. Using the WSe$_2$ heterostructure, we
infer coherence and density dynamics of the exciton 2S state. Owing
to its increased radiative lifetime, at low temperatures, the
dephasing of the 2S state is longer than of the 1S transition. While
scanning various heterostructures across tens of micrometers, we
conclude that the disorder, principally induced by strain
variations, remain to be present creating spatially varying
inhomogeneous broadening.

\end{abstract}


\date{\today}

\maketitle


\section{Introduction} The interest in two-dimensional (2D)
layered semiconductors stems from novel functionalities that can be
obtained by varying their thickness, down to the single-layer (SL)
limit, or by stacking layers from different materials to create van
der Waals heterostructures. Semiconducting transition metal
dichalcogenides (TMDs) in the SL form display direct band
gap\,\cite{MakPRL10} spanning across near infra-red and visible
range. They typically show 10\,$\%$ absorption, which can approach
unity, by engineering their photonic environment\,\cite{Horng19}.
They are thus promising for future flexible optoelectronic
applications, such as: LEDs, lasers, photovoltaics, photodetectors,
electroluminescence or optical modulators\,\cite{MakNatPhot16}.

On the more fundamental level, SL-TMDs are nowadays a benchmark to
study the physics of excitons (EXs), hydrogen-like bound states
formed by Coulomb interaction between an electron and a hole. In
typical direct band gap bulk semiconductors such as GaAs, due to the
large dielectric screening and small effective mass of the carriers,
the resulting EX binding energy is barely a few millielectronvolt
(meV). This is negligible compared to the thermal fluctuations,
unless the sample is studied at low temperatures. In SL-TMDs though,
due to large effective masses and reduced dielectric screening, the
EX binding energy is increased up to several hundred meV, such that
even at room temperature the optical properties are governed by EX
transitions. Furthermore, their band structure displays a pair of
non-equivalent valleys as regards the spin degree of freedom,
establishing original selection rules for the excitation and
emission of light, where the EX valley index is locked with the
light helicity.

SL-TMD have been initially fabricated as bare flakes deposited
directly on SiO$_2$ substrates. In such configuration, they suffer
from a direct contact with the environment and its invasive
chemistry, thus become unstable and prone to structural
aging\,\cite{GaoACSNano16} and
photodegradation\,\cite{AhnACSNano16}. Due to microscopic and
macroscopic disorder factors --- such as: strain, wrinkling, flake
deformations and cracks, lattice defects and vacancies, variation of
charge state and of dielectric constant in the substrate --- the
optical response was largely dominated by the inhomogeneous
broadening\,\cite{MoodyNatCom15, JakubczykNanoLett16}, $\sigma$.
Recently, it was demonstrated that surface protection by
encapsulating SL-TMDs between layers of hexagonal boron nitride
$h$-BN\,\cite{CadizPRX17, Ayayi2DMater17, MancaNatCom17,
JakubczykACSNano19} provides stable samples with low disorder,
avoiding surface contaminations causing the EX broadening.
Encapsulation thus yields step-like improvement of optical
properties. The main indicator of such improvement was the narrowing
of the EX linewidth and appearance of the Rydberg series of the EX's
excited states\,\cite{ChernikovPRL14, MancaNatCom17, MolasPRL19,
GorycaNatCom19}.

\section{Experiment} The question arises if the structural disorder in such
heterostructures can be suppressed to the point, where $\sigma$ is
not detectable in the optical response\,? If so, how can one
demonstrate that the EX transition linewidth indeed attains the
homogeneous limit, $\gamma$\,? Another query regards the spatial
extensions, on which such optimal conditions can be maintained. To
address these issues one has to use an optical microscopy technique
capable to separate $\sigma$ and $\gamma$ within the transition
spectral lineshape. Four-wave mixing (FWM) spectroscopy, which is
the principal tool in our work, has been conceived for this very
purpose.

\begin{figure}[t]
\includegraphics[width=1.03\columnwidth]{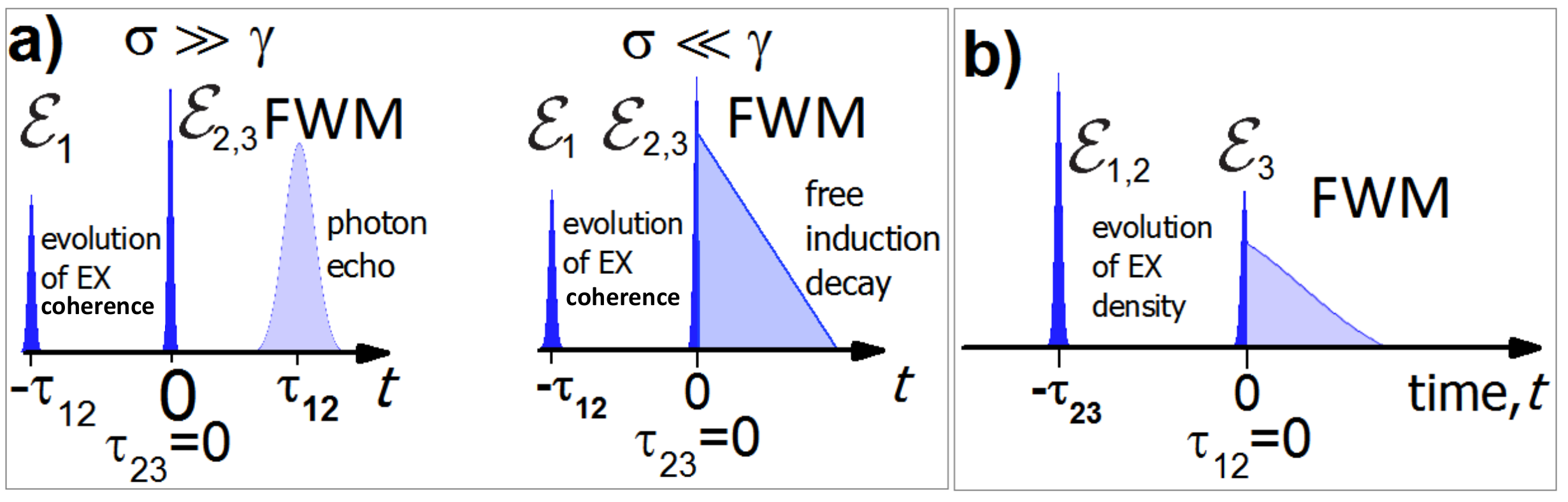}
\caption{{\bf Schema of pulse sequences used in the experiments and
resulting forms of the four-wave mixing transients} a)\,Exciton
coherence dynamics is measured by varying $\tau_{12}$ and keeping
$\tau_{23}=0$. In a presence of inhomogeneous broadening TR-FWM is a
photon echo. When homogeneous broadening dominates, FWM is a free
induction decay. b)\,Exciton density dynamics is probed when varying
$\tau_{23}$ and fixing $\tau_{12}=0$.\label{fig:Fig_Seq}}
\end{figure}

FWM is a nonlinear polarization in the material, proportional to
$\mu^4\Ea^{\star}\Eb\Ec$, where $\Eo$ are three short laser pulses
generated by a femtosecond laser and $\mu$ is the transition
oscillator strength. In this approach, $\sigma$ and $\gamma$ can be
differentiated, by monitoring time-resolved (TR) FWM, while varying
the time delay between $\Ea$ and $\Eb$, $\tau_{12}$. In the presence
of $\sigma$, FWM appears as a photon echo\,\cite{FigS1}. As depicted
in Fig.\,\ref{fig:Fig_Seq}\,a, TR-FWM is a Gaussian peak centered at
$t=\tau_{12}$ and displaying the temporal width proportional to
$1/\sigma$. Such en echo builds up due to the phase conjugation
(labeled with $\star$) between the first order absorption created by
$\Ea$ and FWM. Conversely, in the homogeneous limit, FWM transient
shows the exponential decay starting at $t=0$ for whatever
$\tau_{12}$, also drawn in Fig.\,\ref{fig:Fig_Seq}\,a. This is known
as free induction decay, adopting the nomenclature used in the NMR
spectroscopy. In general, $(\sigma,\,\gamma)$ pair can be accurately
retrieved by a two-dimensional fitting\,\cite{JakubczykACSNano19} of
TR FWM for various $\tau_{12}$.

To enable microscopy, FWM is here detected through optical
heterodyning. In short, i)\,$\Eo$ are modulated by distinct radio
frequencies $\Omega_{1,\,2,\,3}$ around 80\,MHz using acousto-optic
deflectors. $\Eo$ propagate co-linearly and are focussed onto the
sample down to the diffraction limit using a microscope objective of
NA=0.65. ii)\,in the detection path, the reflected signal is
interfered with the reference beam $\Er$ modulated at the
phase-locked linear combination of radio-frequencies
$\Omega_3+\Omega_2-\Omega_1$, sensitive to the FWM response.
iii)\,background-free spectral interference between $\Er$ and FWM is
then recorded as DC signal on a CCD camera, which is installed at
the output of the imaging spectrometer\,\cite{LangbeinOL06}. FWM
field is obtained \emph{via} spectral interferometry, granting
access to the amplitude and phase of the signal. Thus, the modeling
often requires to apply complex response functions, to properly
account for detection of the field and potential interference
effects. This setup is particularly suited to perform FWM
micro-spectroscopy with a high signal to noise ratio, accessing the
EX coherence dynamics with an enhanced temporal (spatial) resolution
of 150\,fs (500\,nm). Further details regarding the experimental
technique can be found in Refs.\,[\onlinecite{LangbeinOL06,
LangbeinRNC10}], while sketches of the experimental setup are
presented in supplementary material in
Refs.\,[\onlinecite{FrasNatPhot16, Jakubczyk2DMat17}].

\section{Results}

\subsection{Photoluminescence and FWM mapping of a MoSe$_2$ heterostructure}
\begin{figure}[t]
\includegraphics[width=1.03\columnwidth]{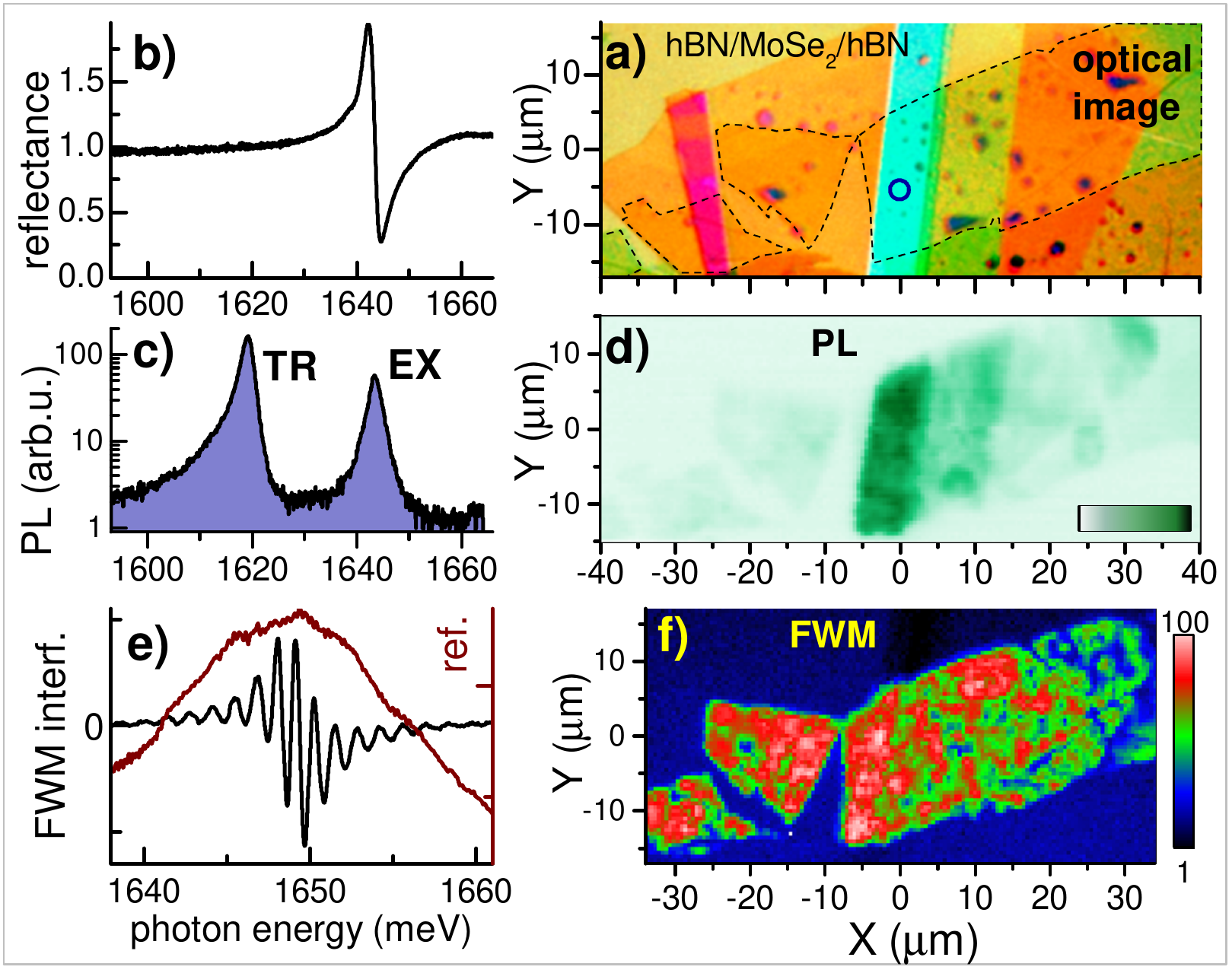}
\caption{{\bf Characterization of a $h$-BN/MoSe$_2$/$h$-BN
heterostructure \emph{via} optical microspectroscopy, T=5\,K.}
a)\,Optical image of the sample. Different colors correspond to
different thicknesses of the bottom $h$-BN layer. b)\,Measured
micro-reflectance from the area marked with a blue circle in a),
showing EX linear response. c)\,PL spectrum measured at the same
location as in b) showing neutral (EX) and charged exciton (TR)
transitions. d)\,Imaging of the spectrally integrated PL.
e)\,Typical spectral interference (black) between FWM at zero delay
and $\Er$ (brown trace). f)\,Imaging of the time-integrated FWM
amplitude. \label{fig:Fig1}}
\end{figure}

We first present results obtained on a $h$-BN/MoSe$_2$/$h$-BN
heterostructure. Its optical image under white light illumination,
is shown in Fig.\,\ref{fig:Fig1}\,a. The edges of the MoSe$_2$ flake
are contoured with the dashed line. The dark spots correspond to the
air blisters between the flake and the top $h$-BN having the
thickness around 20\,nm. The bottom $h$-BN instead shows terraces of
different thickness 232\,nm (blue area), 240\,nm (yellowish),
252\,nm (orange) and 290\,nm (red). To assess the optical response,
the heterostructure is cooled down to T=5\,K and we perform confocal
photoluminescence (PL) under non-resonant CW excitation at 710\,nm,
at the position marked with a blue circle in a). As shown in
Fig.\,\ref{fig:Fig1}\,c, in the PL not only the neutral EX
transition is observed, but also the trion (TR) one. For the latter,
we note the characteristic low energy tail in the PL signal. While,
owing to the residual strain, the spectral position of both
transitions fluctuate across the sample by around 15\,meV, the TR
binding energy is constant and equal to 26\,meV. In
Fig.\,\ref{fig:Fig1}\,d, we present the PL imaging of the sample
obtained \emph{via} raster scanning the excitation and spectrally
integrated over TR and EX transitions. It is interesting to notice
the correlation between the spectrally integrated PL intensity and
the thickness of the bottom $h$-BN terraces: the thinner the latter,
the stronger PL is observed. This result is consistent with the
recent observation that the EX radiative lifetime can be tuned
through the Purcell effect, when varying the thickness of the
surrounding $h$-BN layers\,\cite{FangPRL19}. To investigate this
scenario we propose to perform time-resolved resonant spectroscopy.

To measure the resonant optical response we first perform white
light reflectance restricting the signal to the spot marked with a
blue circle. The result is shown in Fig.\,\ref{fig:Fig1}\,b. Only EX
transition appears as a dispersive quasi-Lorentzian line, with the
full width at half-maximum (FWHM) of marely $2.5\,$meV. The trion
line is absent owing to encapsulation and resonant excitation, which
both suppress density of free carriers required to form charged
excitons. In order to measure coherent nonlinear response FWM
microscopy is carried out. In Fig.\,\ref{fig:Fig1}\,e we present a
typical spectral interferogram between $\Er$ (brown line) and the
FWM signal at $\tau_{12}=\tau_{23}=0$. FWM image is constructed by
plotting time-integrated (TI) FWM amplitude measured for different
points on the sample and is shown in Fig.\,\ref{fig:Fig1}\,f.
Although the strongest FWM is measured at the flake area covered
with the ``blue" $h$-BN stripe, globally there is no clear
correlation between the FWM amplitude, reflecting to large extend
the EX radiative rate, and the thickness of the encapsulating
layers. This is attributed to the underlying disorder, generating
$\sigma$. We have recently demonstrated\,\cite{JakubczykACSNano19}
that there exists correlation between $\sigma$ and the EX lifetime:
the latter is increased at the areas showing larger $\sigma$ due to
the stronger exciton localization. The same effect is present here,
as evidenced in the Supplementary Fig.\,S1. We also note that
different interference conditions from the various thicknesses of
the encapsulating layers generate different electric field
distributions of the excitation laser at a plane of MoSe$_2$ SL.
This could yield different PL intensities measured at different
$h$-BN thicknesses in Fig.\,\ref{fig:Fig1}\,d.

\subsection{Dephasing mechanisms of excitons in the MoSe$_2$ heterostructure}
\begin{figure}[t]
\includegraphics[width=1.03\columnwidth]{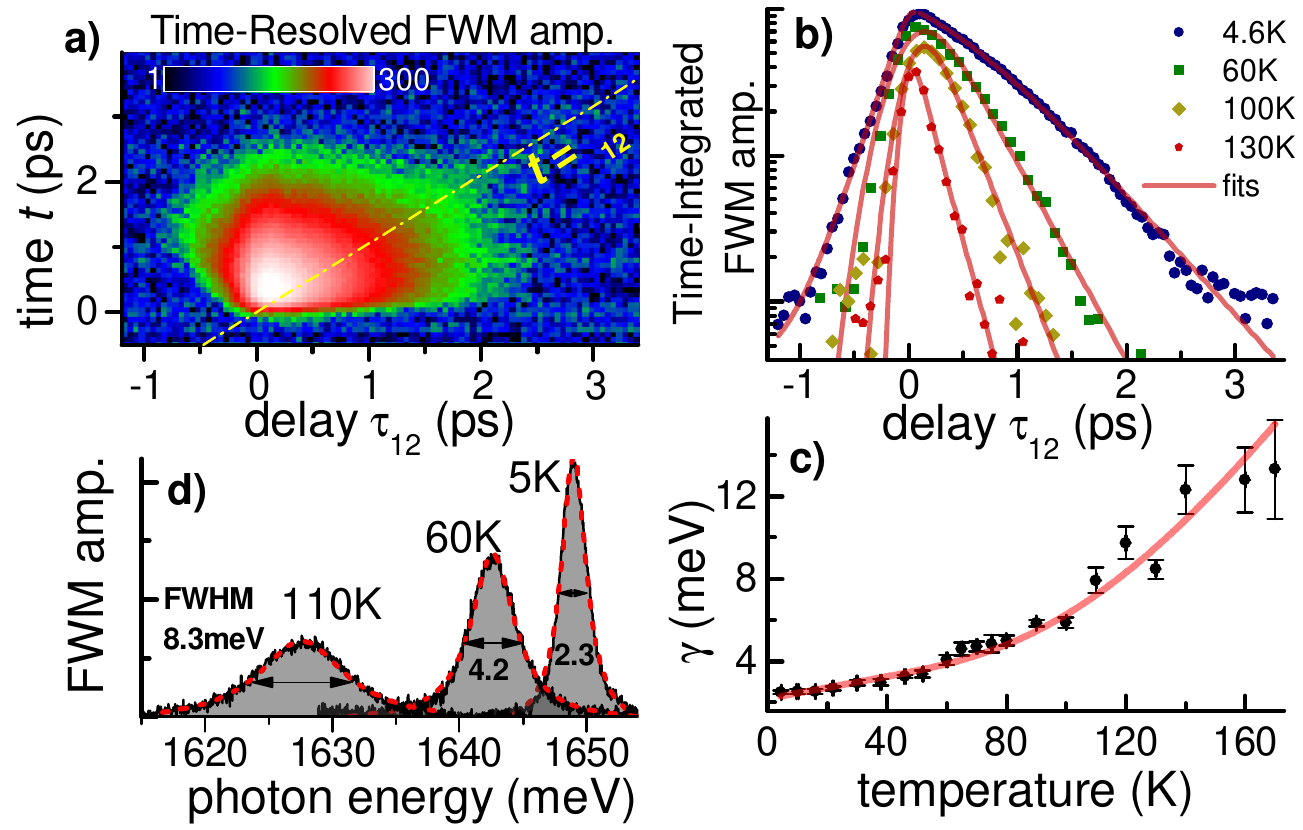}
\caption{{\bf Temperature dependent EX dephasing measured at the
area of MoSe$_2$ heterostructure dominated by the homogeneous
broadening.} a)\,Time-resolved FWM amplitude measured for various
delays $\tau_{12}$. FWM transient appears as free-induction decay
indicating that the response is dominated by $\gamma$.
b)\,Time-integrated FWM amplitudes \emph{versus} $\tau_{12}$
presented for several temperatures along with the fits (red lines).
c)\,Homogeneous broadening $\gamma$ as a function of temperature,
indicating phonon-induced dephasing mechanism, as modeled with the
red line. d)\,Exemplary spectra of FWM amplitude for different
temperatures, $\tau_{12}=0.6\,$ps.\label{fig:Fig2}}
\end{figure}

It was recently reported that FWM transient in bare SL TMDs is a
photon echo\,\cite{MoodyNatCom15, JakubczykNanoLett16,
Jakubczyk2DMat17}. In Fig.\,\ref{fig:Fig2}\,a we present TR FWM for
different delays $\tau_{12}$ measured at the same area as in
Fig.\,\ref{fig:Fig1}\,b. Here, there is virtually no sign of photon
echo formation; the maximum amplitude of the transient does not
follow the diagonal $t=\tau_{12}$. Instead, for $\tau_{12}>0$ FWM is
maximal for $t=0$, and then evolves as the exponential free
induction decay for $t>0$. This shows that the optical response is
dominated by the homogeneous broadening, \emph{i.e.},
$\gamma\gg\sigma$. It is worth to note that FWM is here also
detected for $\tau_{12}<0$. In such configuration, $\Ea$ arrives
last and triggers FWM through the local field
effect\,\cite{WegenerPRA90, JakubczykACSNano19}. Furthermore, as
$\Ea$-$\Er$ delay increases, FWM sets off for increasingly longer
times when going into more negative delays, as indeed visible in the
color map.

To measure $\gamma$, we plot TI FWM \emph{versus} $\tau_{12}$, as
shown in Fig.\,\ref{fig:Fig2}\,b. The data are modeled by a
convolution of a Gaussian, reflecting pulse duration of around
150\,fs and exponential decay $\exp(-\gamma\tau_{12}/2\hbar)$,
parameterized with $\gamma$ (FWHM) or dephasing time
$T_2=2\hbar/\gamma$. At low temperature T=4.6\,K we measure
$\gamma=(2.5\,\pm\,0.1)\,$meV. This differs from around 1\,meV
measured in bare MoSe$_2$ SLs\,\cite{JakubczykNanoLett16} and is
therein attributed to longer EX lifetime due to localization on
disorder. The latter is significantly larger in bare SLs, as
monitored by $\sigma$.

On top of ideal TMD crystal properties --- determining the
morphology of the EX wave function, its binding energy and radiative
lifetime --- there are at least three major environmental parameters
influencing $\gamma$: temperature, EX density and local disorder,
which will be considered in the following.

In Fig.\,\ref{fig:Fig2}\,c we present $\gamma$ measured for
different temperatures. The apparent increase of $\gamma$  is
modeled by the sum of linear increase and exponential activation
part, due to interaction with acoustic and optical phonons,
respectively. The data are fitted with the following equation:
$\gamma({\rm T})=\gamma_0+a{\rm T}+b/[\exp(E_1/k_{B}{\rm T})-1]$,
with $E_1=37\,$meV, corresponding to activation energy of optical
phonons\,\cite{MolinaPRB11}. Qualitatively, the phonon dephasing
mechanisms are therefore similar as in bare MoSe$_2$
SLs\,\cite{JakubczykNanoLett16}. We note that temperature induced
increase of the linewidth can be directly monitored in the FWM
spectra, as exemplified in Fig.\,\ref{fig:Fig2}\,d. Owing to
negligible $\sigma$, there is a very good correspondence between
$\gamma$ retrieved from the coherence dynamics and directly measured
FWM spectral linewidth. We emphasize that such comparison to be
valid, requires prior knowledge that the transition lineshape is
homogeneous. Such an independent assessment is here provided by TR
FWM data.

\begin{figure}[t]
\includegraphics[width=1.05\columnwidth]{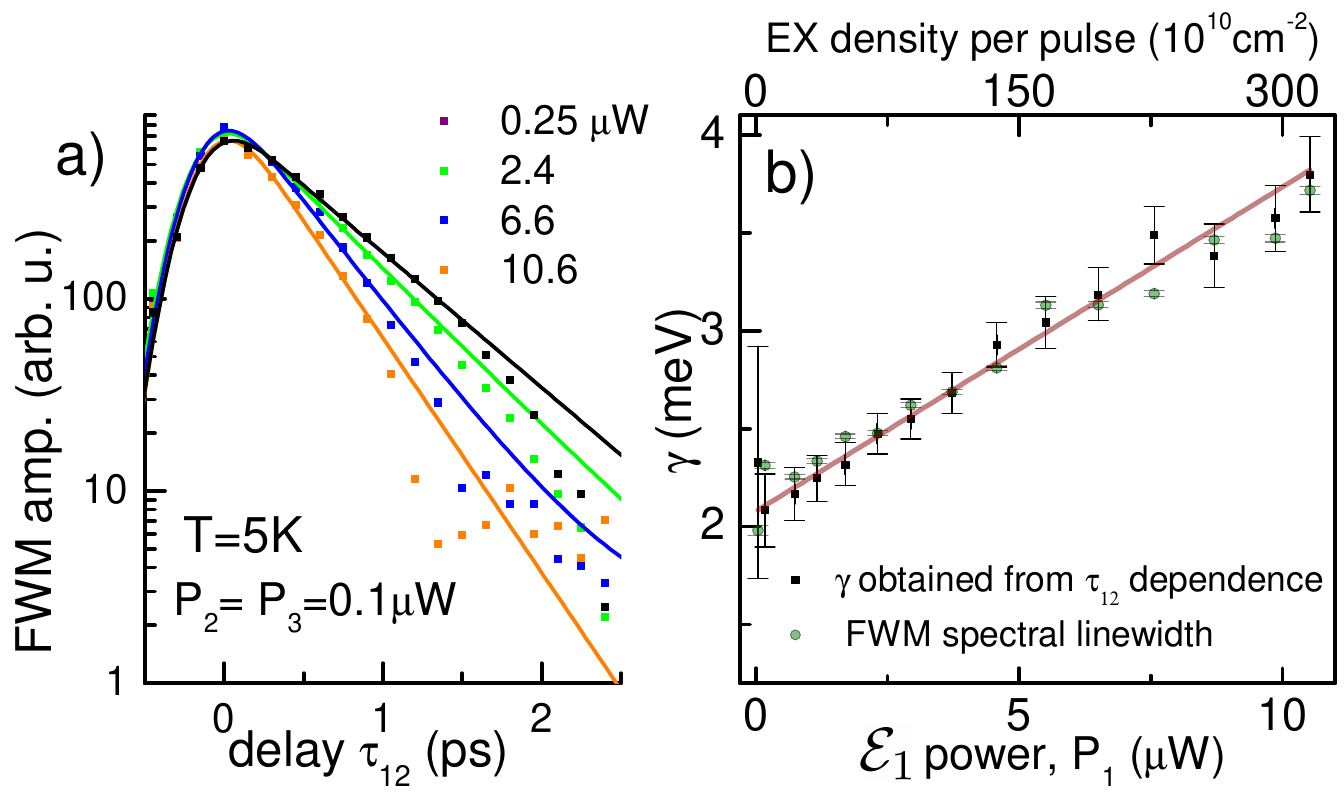}
\caption{{\bf EX dephasing measured as a function of the resonantly
generated density at T=5\,K at the area of MoSe$_2$ heterostructure
entirely dominated by homogeneous broadening.}
a)\,$\tau_{12}$-dependence of the time-integrated FWM amplitude
measured for different $\Ea$ excitation powers, $P_1$.
b)\,Homogeneous broadening $\gamma$ versus EX density (black
squares), compared with FWM spectral FWHM (green circles).
\label{fig:Fig3}}
\end{figure}

An important factor influencing $\gamma$ is the EX density, $n$.
With its increase, EX-EX interaction increases, causing excitation
induced dephasing, as observed in the past in semiconductor quantum
wells\,\cite{WangPRL93, HonoldPRB89} and more recently for
WSe$_2$\,\cite{MoodyNatCom15} and MoS$_2$\,\cite{JakubczykACSNano19}
SLs. In Fig.\,\ref{fig:Fig3}\,a we present exemplary curves of
coherence dynamics, where a steeper slope is visible with increasing
$\Ea$ intensity, $P_1$. Monitoring the impinging power and assuming
absorption of approximately 10\%, we can estimate $n$. In
Fig.\,\ref{fig:Fig3}\,b we plot the resulting $\gamma(n)$
dependence: the broadening is well visible, we observe that $\gamma$
can be doubled with 300-fold increase of the resonantly generated EX
density. We propose to describe the effect by the linear dependence
$\gamma(n)=\gamma_0+\alpha\times n$, where $\gamma_0$ is the
zero-density linewidth and $\alpha$ is the interaction parameter,
both established for T=5\,K. From the linear fit, we obtain
$\gamma_0=(2.07\,\pm\,0.03)\,$meV and
$\alpha=(5.4\,\pm\,0.2)\times10^{-13}\,$meVcm$^{2}$. The interaction
strength is here 5 times smaller compared to similar studies
performed on WSe$_2$ SLs\,\cite{MoodyNatCom15}. In
Fig.\,\ref{fig:Fig3}\,b we also show that the broadening is directly
visible in the FWM spectra and in a good agreement with the
coherence dynamics measurements.

\subsection{Reflectance and FWM response in a WSe$_2$ heterostructure}

\begin{figure}[t]
\includegraphics[width=1.03\columnwidth]{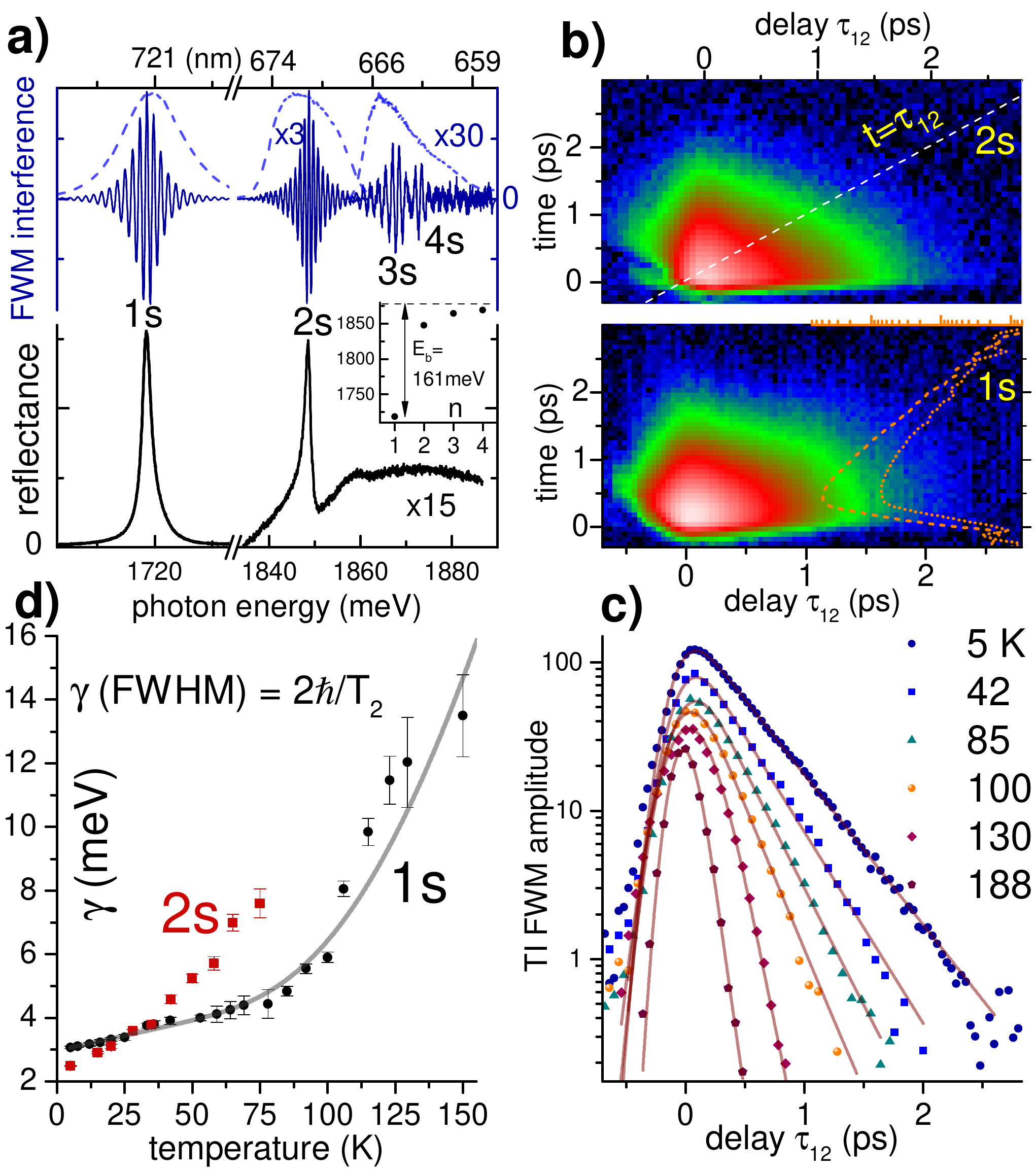}
\caption{{\bf Four-wave mixing microspectroscopy of the exciton
transitions in a WSe$_2$ heterostructure.} a)\,Measured
micro-reflectance showing linear response of 1s and 2s states
(bottom) and FWM-reference spectral interference revealing exciton
nS series up to n=4. Inset: energy \emph{versus} quantum number n.
b)\,Transients of the FWM amplitude versus $\tau_{12}$ showing
free-induction decay (directly exemplified with orange traces for
$\tau_{12}=0.2\,$ps and 1\,ps) for 1S and 2S states.
c)\,Time-integrated FWM amplitudes as a function of $\tau_{12}$ for
temperatures as indicated. d)\,Temperature dependence of the
homogeneous broadening $\gamma$ for 1S and 2S states. The modeling
is indicated with a gray trace.\label{fig:Fig4}}
\end{figure}

The second part of this work is devoted to investigations of a
WSe$_2$ heterostructure. Its optical image is presented in
Fig.\,\ref{fig:Fig6}\,a. A large WSe$_2$ SL of around
$(60\,\times\,60)\,\mu$m was transferred at the bottom $h$-BN
$(90\,\pm\,10)\,$nm thick layer and then partially covered with the
top one of nominally the same thickness, resulting in the elongated
heterostructure, with a shape marked with a dashed contour in
Fig.\,\ref{fig:Fig6}\,a. Let us first investigate the area of the
flake marked with a cross, which is free from wrinkling (occurring
at the edges) and air blisters (visible as dark spots on the
neighboring surface).

The reflectance spectrum measured at T=5\,K is presented in
Fig.\,\ref{fig:Fig4}\,a. The fundamental EX 1S transition is
detected as an emissive quasi-Lorentzian line, with FWHM only
$2.6\,$meV. Interestingly, the second resonance is clearly observed
around $130\,$meV above EX 1S, with even a sharper line-shape of
$1.7\,$meV (FWHM). It is attributed to the EX excited 2S state. An
additional spectral wiggle is seen another $10\,$meV higher, however
its contrast is not sufficient for the unequivocal assignment of the
underlying transitions. FWM spectral interferograms at $\tau_{12}=0$
are shown as blue traces in Fig.\,\ref{fig:Fig4}\,a. To drive FWM of
the 1S EX we still employ a Ti:Shapphire laser. However, to study
the excited states occurring well below 700\,nm, an Optical
Parametric Oscillator must be used. As the technique is background
free and detection is at the shot-noise limit, apart from 1S and 2S
states, FWM reveals further resonances attributed to 3S and 4S
states. When plotting the energy versus the quantum number (see
inset) we observe non-hydrogenic series, in line with previous
findings\,\cite{ChernikovPRL14} and the EX binding
energy\cite{MolasPRL19, GorycaNatCom19} of E$_{\rm b}$=161\,meV.

\subsection{Exciton coherence dynamics in the WSe$_2$ heterostructue:
dephasing and coherent coupling} FWM transients \emph{versus}
$\tau_{12}$ measured for 1S and 2S states are shown as color maps in
Fig.\,\ref{fig:Fig4}\,b. With increasing $\tau_{12}$, FWM is emitted
as an exponentially decaying transient with a maximum centered
around $t=0$. We see strictly no indication of the photon echo
formation. FWM takes the form of the free induction decay, proving
that at the probed position, the EX transition is homogeneously
broadened. To determine FWHM of $\gamma=2\hbar/T_2$, in
Fig.\,\ref{fig:Fig4}\,c we plot time-integrated FWM amplitude as a
function of $\tau_{12}$. Resulting coherence dynamics displays a
mono-exponential decay over three (six) orders of magnitude in a FWM
field (intensity). From these decays we extract $\gamma$, similarly
as for the MoSe$_2$ heterostructure (Fig.\,\ref{fig:Fig2}\,b).

\begin{figure}[t]
\includegraphics[width=1.05\columnwidth]{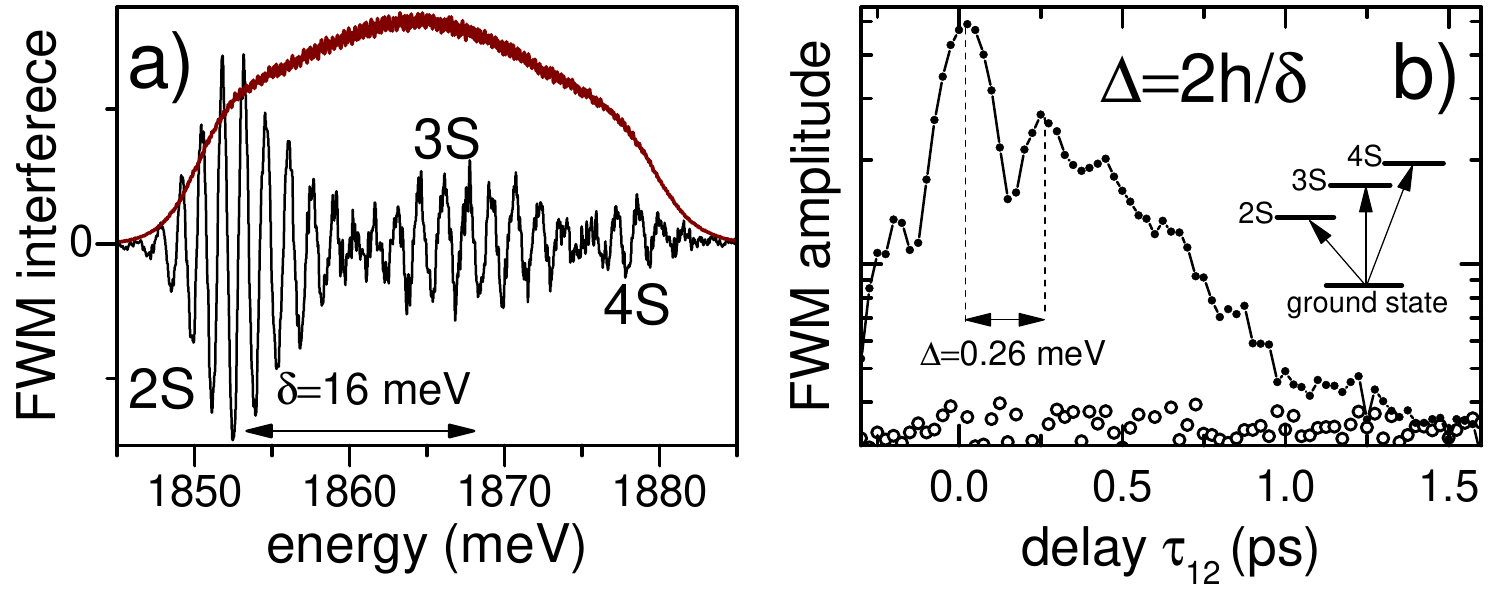}
\caption{{\bf Coherent coupling in the EX excited states measured in
the WSe$_2$ heterostructure.} a)\,FWM spectral interferogram showing
EX excited states 2S, 3S and 4S (black), simultaneously excited by
the OPO laser pulse (brown). b)\,The beating measured in the
coherence dynamics \emph{via} TI FWM$(\tau_{12})$ indicates the
presence of coherent coupling between the excites states, especially
between 2S and 3S, which dominate the spectrum. \label{fig:Fig4a}}
\end{figure}

Let us note that the excitation bandwidth of the OPO laser is
sufficient to simultaneously cover the excited states 2S, 3S and 4S,
as shown in Fig.\,\ref{fig:Fig4a}\,a. Because these transitions
share the same ground state, they are expected to be coherently
coupled (absorption on one of the transitions, mutually generates
FWM on the other ones). Such coupling should therefore induce
quantum beating in the corresponding coherence dynamics. In
Fig.\,\ref{fig:Fig4a}\,b, we present such dynamics measured
\emph{via} $\tau_{12}$-dependence. Indeed, we observe that FWM is
here modulated with the period of
$\Delta=2\pi\hbar/\delta=0.26\,$meV, well corresponding to the 2S-3S
splitting $\delta\simeq16\,$meV. We can resolve two oscillation
periods within the initial $0<\tau_{12}<0.5\,$ps. For subsequent
delays $\tau_{12}>0.5\,$ps the beating is smeared out owing to fast
dephasing of higher excited states. This measurement thus indicates
that one can induce couplings within the Rydberg states of TMD EXs.

Focussing back on the dephasing mechanisms, in
Fig.\,\ref{fig:Fig4}\,c we observe that for increasing temperature,
the FWM decay gets more pronounced, owing to the increase of
$\gamma$ (and correspondingly, shortening of dephasing time $T_2$).
Such temperature dependence of $\gamma$, measured at 1S and 2S
states\,\cite{FigS2} is summarized in Fig.\,\ref{fig:Fig4}\,d.

The exciton 1S EX transition is once again modeled with a linear
increase, with $\gamma_0=(2.96\,\pm\,0.05)\,$meV and
$a=(0.02\,\pm\,0.002)\,$meV/K, attributed to the interaction with
acoustic phonons, and the exponential activation term with
$E_1=44\,$meV, due to the onset of interaction with optical
phonons\,\cite{MolinaPRB11}.

As regards the 2S EX transition, let us remind that for an ideal
two-dimensional Wannier exciton, the oscillator strength $\mu$
scales with the quantum number $n$ as: $\mu\propto(n-1/2)^{-3}$. The
radiatively limited broadening of the 2S EX could be therefore
estimated to be $3^3=27$ times weaker than the one of the 1S EX,
thus only around $0.1\,$meV at low temperature. It is interesting to
note that for $T<20\,K$, we indeed measure $\gamma(2S)<\gamma(1S)$.
However $\gamma(2S)$ is around 20 times larger than the radiative
limit, indicating that even low temperature dephasing is dominated
by a non-radiative decay, plausibly through the scattering into 1S
EX at high center-of-mass in-plane momenta. With increasing
temperature, we observe much stronger EX dephasing for the 2S state
with respect to 1S. However, recent theoretical
work\,\cite{BremNanoscale19} predicts that 2S transition is less
prone to interaction with phonons than the 1S. We here therefore
refrain from the modeling of the $\gamma(T)$ dependence measured for
the 2S EX.

\subsection{Population dynamics of 1S and 2S excitons}
\begin{figure}[t]
\includegraphics[width=1.03\columnwidth]{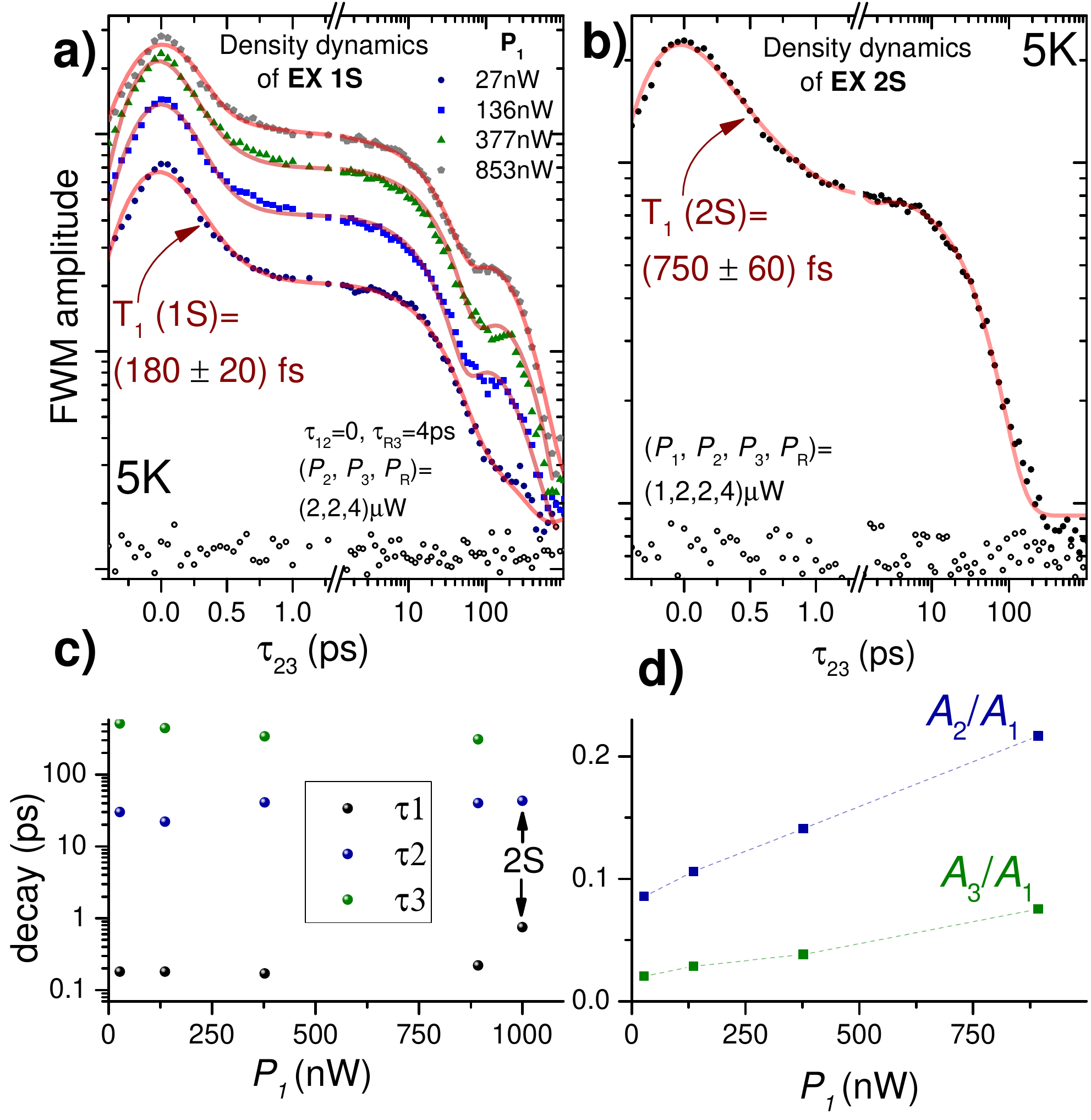}
\caption{{\bf Time-integrated FWM as a function of delay $\tau_{23}$
reflecting resonantly excited EX density dynamics in the WSe$_2$
heterostructure} a)\,Density dynamics of EX 1S measured for
different $\Ea$ excitation powers, as indicated. Modeling is given
with red traces. b)\,as in a), but measured for EX 2S. Longer
radiative decay of the 2S state with respect to the 1S has its
impact on the initial population loss, which is measured to be more
than four times longer for the 2S state. Both measurements were
performed at the same location of the flake. c)\,Decay constant
$\tau_1$, $\tau_2$ and $\tau_3$ retrieved from the modeling of 1S EX
density (a) and 2S EX density dynamics (b). d)\,Amplitude ratios
$A_2/A_1$ and $A_3/A_1$ \emph{versus} $P_1$ obtained from the
modeling.\label{fig:Fig5}}
\end{figure}

To further elucidate the interplay between radiative and
non-radiative decay mechanisms of EX 1S and 2S transitions, we infer
the dynamics of resonantly excited EX density. We set $\tau_{12}=0$
and employ $\Ea$ and $\Eb$ to drive modulation of EX's population,
slowly varying with the frequency $\Omega_2-\Omega_1=1\,$MHz.
TI-FWM, triggered \emph{via} $\Ec$ delayed from $\Eb$ by
$\tau_{23}$, reflects the amplitude of this oscillating EX density,
as depiced in Fig.\,\ref{fig:Fig_Seq}\,b. Exemplary data reflecting
such dynamics of 1S EX measured at T=5\,K for increasing $P_1$ and
fixed $P_2$ and $P_3$ are given in Fig.\,\ref{fig:Fig5}\,a. While
the EX coherence, inspected in previous sections, vanishes after a
few ps, the EX density evolves here on a timescale of at least one
nanosecond. This is consistent with recent FWM experiments on SL
MoSe$_2$\,\cite{ScarpelliPRB17} and MoS$_2$
heterostructure\,\cite{JakubczykACSNano19}, and also perovskite
nanocrystals\,\cite{BeckerNanoLett18} and
nanoplatlets\,\cite{NaeemPRB15}.

Qualitatively, in the dynamics we distinguish: i)\,the rise from
negative $\tau_{23}$, owing to the finite duration of the laser
pulses, ii)\,initial decay on a few hundred fs scale, attributed to
the radiative decay and EX scattering to dark EX states.
iii)\,followed after $\tau_{23}>1\,$ps by a much longer FWM decay,
due to populating the dark EX states and their subsequent relaxation
and interaction with the bright ones, \emph{i.e.}, EXs within the
light cone, obeying spin selection rules for radiative
recombination.

Because the density dynamics is here measured \emph{via} the
interference with $\Er$, which is phase-sensitive, it should be
described with a complex response function $R$. The latter is here
proposed in a form of a coherent superposition of three exponential
decay components\,\cite{ScarpelliPRB17, JakubczykACSNano19}:

\begin{align}
R(\tau_{23})=\sum_{j=1}^{3}A_j\Theta(\tau_{23})\exp({i\phi_j-\tau_{23}/\tau_j})\label{Eq:ComplexFit}
\end{align}

where $(A_j,\,\phi_j,\,\tau_j)$ with $j={(1,\,2,\,3)}$ are
amplitude, phase and decay constant of each of the three processes.
The pulse width is taken into account by convoluting $R$ with the
Gaussian of $\tau_0=150\,$fs. Furthermore, owing to the repetition
period $T_R\simeq13\,$ns, there is a pile-up effect building up
throughout many repetitions of the heterodyne experiment equal to
$(\exp({\frac{T_R}{\tau_j}})-1)^{-1}$. The response function
therefore takes the following form:
$R(\tau_{23})=\sum_{j=1}^{3}A_j\{[\exp({\frac{T_R}{\tau_j}})-1]^{-1}+\frac{1}{2}[(1+{\rm
erf}(\frac{\tau_{23}}{\tau_0}-\frac{\tau_{0}}{2\tau_j})]\}\times\exp{(i\phi_j-\frac{\tau_{23}}{\tau_j}+\frac{\tau_0^2}{4\tau_j^2})}$.
Finally, to fit the FWM amplitude we take absolute value of
$R(\tau_{23})$.

The result of the modeling is given by red traces and retrieved set
of decay constants for different $P_1$ is given in
Fig.\,\ref{fig:Fig5}\,c. For the initial decay we consistently find
$\tau_1=T_1({\rm 1S})\simeq0.18\,$ps. The second and third process
decay on a scale of a few tens and a few hundred ps, respectively.
They describe EX evolution in the dark state reservoirs, namely EXs
with the in-plane momentum out of the light cone and the
spin-forbidden ones. Note that the wiggling observed around
$\tau_{23}=100\,$ps, building up when increasing $P_1$, is due to
interference between all three processes. The interference becomes
more pronounced as the ratios $A_2/A_1$ and $A_3/A_1$ increase with
$P_1$, as reported in Fig.\,\ref{fig:Fig4a}\,d. In other words,
increasing the excitation results in higher occupancy of dark EX
states and their enhanced interaction with the bright EX, which is
probed with FWM $\tau_{23}$-dependence.

In Fig.\,\ref{fig:Fig5}\,b we present the density dynamics measured
on the 2S EX transition. The modeling includes now only two decay
processes. Interestingly the initial decay of $\tau_1=T_1({\rm
2S})\simeq0.75\,$ps is measured to be more than four times longer
than for the 1S EX, indicating longer radiative decay of the 2S
state. It is however still far off from the prediction based on the
relative oscillator strength, again indicating dominance of the
non-radiative decay processes of the 2S states. We note that
dephasing of the higher states is much faster than our temporal
resolution of 150\,fs, showing prominently non-radiative decay of
the higher excited EX states.

\subsection{FWM mapping of exciton 1S and 2S transitions in the WSe$_2$ heterostructure}
\begin{figure}[t]
\includegraphics[width=\columnwidth]{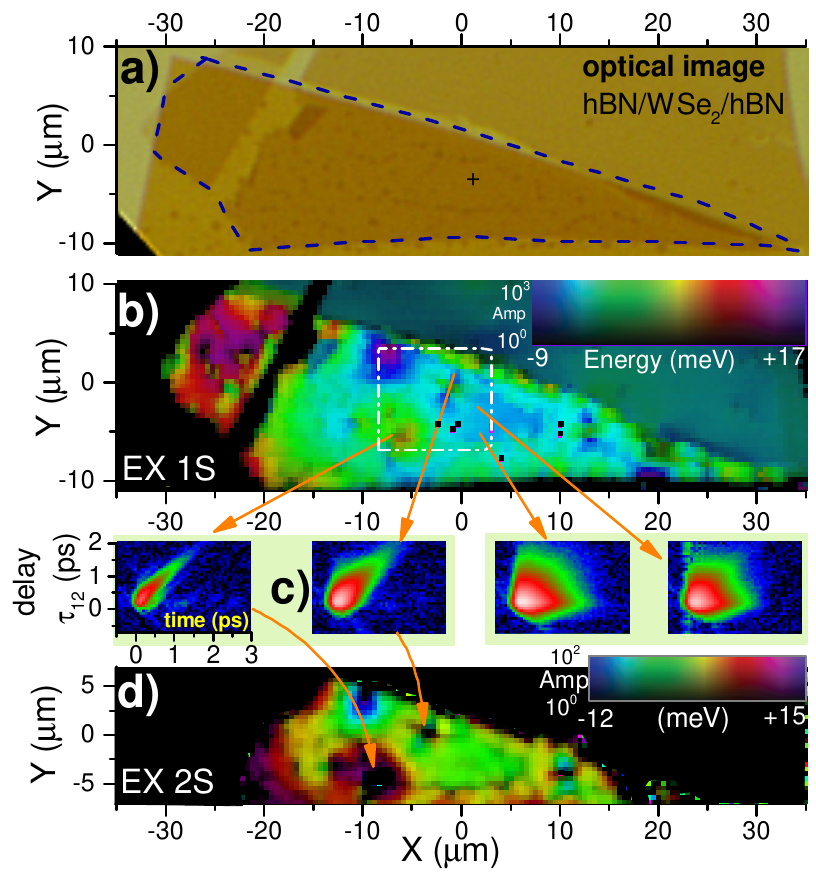}
\caption{{\bf Four-wave mixing imaging of 1S and 2S EXs in the
WSe$_2$ heterostructure.} a)\,Optical image of the sample. The
heterostructure area is enclosed with a dashed line. b)\,FWM imaging
of the 1S EX, $\tau_{12}=0.1\,$ps. Amplitude and transition energy
(with respect to central energy of 1720\,meV) are encoded as
brightness and hue level, as indicated. Note that WSe$_2$ not
covered with the $h$-BN also generates FWM signal. c)\,Exemplary
time-resolved FWM \emph{versus} $\tau_{12}$. The areas displaying
strong energy gradients, owing to varying strain, generate photon
echo. Free induction decay is measured on the locations where the
transition energy is quasi-constant. d)\,FWM imaging of the 2S EX,
$\tau_{12}=0.1\,$ps. Scale as in b), central energy is
1845\,meV.\label{fig:Fig6}}
\end{figure}

An important conclusion emerging from our experiments is that EX's
excited states can be observed uniquely at the areas with eradicated
disorder, where the optical response is dominated by $\gamma$. To
substantiate this claim, we perform FWM spatial imaging across the
flake. In Fig.\,\ref{fig:Fig6}\,b we present a color map where FWM
amplitude and transition energy of 1S EX are encoded as brightness
and hue level, respectively. Even though locally (on the areas up to
$(5\,\times5)\,\mu$m$^2$) the transition energy can be
quasi-constant (spectrally fluctuating an order of magnitude less
than $\gamma$), globally the EX energy changes by around 25\,meV.
This variation is attributed to residual strain generated during the
processing of the heterostructure - other sources of disorder, like
variation of dielectric permittivity or doping were shown to be
suppressed\,\cite{RajaNatNano19} in $h$-BN encapsulated TMD SLs.

To illustrate the interplay between strain-induced disorder and
coherent response, we perform mapping of the EX coherence dynamics:
we therefore spatially scan within the part of the heterostructure
enclosed by a white square, while at each position the delay
$\tau_{12}$ is varied. In Fig.\,\ref{fig:Fig6}\,c we provide four
examples of the time- and $\tau_{12}$-resolved FWM. In the first
pair of panels, the response clearly displays the photon echo. When
cross-correlating with b), one assigns these to the areas exhibiting
a strong radial strain gradient of the EX energy, generating
$\sigma$. Conversely, the second pair of panels, measured on a spot
showing a quasi-constant energy level, yields FWM as a
free-induction decay signal. We note that at some positions, we
observe FWM temporal response generated by a mixture of
inhomogeneous contributions. In other words, even within the
sub-$\mu$m excitation area, several domains characterized by
different $\gamma$ and $\sigma$ can contribute to the signal,
rendering quantitative analysis of these transients challenging.

Finally, we have carried out FWM spatial imaging of the 2S EX
transition. As in panel b), we encode the amplitude and transition
energy as brightness and hue, and present the result in
Fig.\,\ref{fig:Fig6}\,d. One can notice that at areas with
suppressed disorder --- here identified by FWM displaying
free-induction decay instead of photon echo ---  the 2S EX
transition can be observed. Interestingly, FWM of the 2S EX is
suppressed at the areas exhibiting a pronounced photon echo. This
indicates that absorption of EX's excited states can only be induced
at locations with low disorder on both microscopic and macroscopic
scales.

\section{Conclusions} We investigated coherent non-linear
response of the EX transitions in high quality
$h$-BN/MoSe$_2$/$h$-BN and $h$-BN/WSe$_2$/$h$-BN heterostructures.
Using FWM microscopy and imaging we showed that encapsulating SLs of
MoSe$_2$ and WSe$_2$ generally provides low disorder samples, which
on areas reaching $(5\,\times\,5)\,\mu{\rm m}^2$ host homogeneously
broadened EX transitions. At such locations, we: i)\,investigated EX
dephasing mechanisms, ii)\,revealed coherent nonlinear response of
EX excited states up to 4S, iii)\,provided indication of coherent
coupling within excited states. We compared population lifetimes of
1S and 2S states, notably finding longer dephasing time (smaller
homogeneous broadening) of the 2S state. This is assigned to its
larger spatial extent and correspondingly longer radiative lifetime
with respect to the EX 1S, as indicated by investigation of the EX
population dynamics. Even though locally optical response of these
TMD heterostructures is nowadays brought close to perfection, our
work shows that further efforts in processing are necessary, to
attain such performance across the entire heterostructure.
Eliminating spatially varying disorder would, for instance, permit
to quantitatively evaluate the impact of dielectric surrounding on
the coherent response of TMD EXs. Furthermore, the resulting weak EX
localization would enable to investigate propagative aspects of EX
coherence and density (diffusion) by implementing FWM with a
spatially separated excitation beams.

\section{Methods: sample fabrication}
The heterostructures were obtained by a bottom-up method using the
gel-film assisted exfoliation technique. As a substrate an
ultra-flat silicon wafers were used, provided by Ted Pella. The
WSe$_2$ and MoSe$_2$ crystals were bought from HQ Graphene and the
high purity $h$-BN was prepared in NIST (Japan). A bottom layer of
$h$-BN was exfoliated directly to the silicon substrate. The flakes
of required thickness were identified using the optical microscope
and some of them verified using AFM. For the heterostructures
$h$-BN/WSe$_2$/$h$-BN the thickness of bottom $h$-BN is
$(90\,\pm\,10)\,$nm. The heterostructures $h$-BN/MoSe$_2$/$h$-BN
possess sequence of different thicknesses in the range from 200 to
250\,nm for the bottom $h$-BN. SLs of TMDs and top $h$-BN were
transferred directly from gel-film using all-dry transfer method.
Heterostructures were annealed before each transfer
(180$^{\circ}\,$C, 5\,min) to remove as much residues as possible.
After completing the stacking process all samples were annealed for
another 5 hours at 200$^{\circ}\,$C. All heterostructures contain
some air trapped between TMD and top $h$-BN flake (easily visible in
form of bubbles in Fig.\,\ref{fig:Fig1}\,a and
Fig.\,\ref{fig:Fig6}\,a).

\section{Acknowledgments} JK acknowledges support by the European
Research Council (ERC) Starting Grant PICSEN (grant no. 306387) and
he is grateful for discussions with D.~Wigger, M.~Richter, M.~Selig
and J.~Renard and W.~Langbein. LZ acknowledges the support from the
Student Project IGA\,Pr\,F2019\,031 of Palack\'{y}. MB acknowledges
the support from the Ministry of Education, Youth and Sports of the
Czech Republic under the project CEITEC 2020 (LQ1601) University.
Growth of hexagonal boron nitride crystals was supported by the
Elemental Strategy Initiative conducted by the MEXT, Japan and the
CREST (JPMJCR15F3), JST. Finally, we acknowledge support by the EU
Graphene Flagship project (no. 785219).


\begin{thebibliography}{30}%
\makeatletter
\providecommand \@ifxundefined [1]{%
 \@ifx{#1\undefined}
}%
\providecommand \@ifnum [1]{%
 \ifnum #1\expandafter \@firstoftwo
 \else \expandafter \@secondoftwo
 \fi
}%
\providecommand \@ifx [1]{%
 \ifx #1\expandafter \@firstoftwo
 \else \expandafter \@secondoftwo
 \fi
}%
\providecommand \natexlab [1]{#1}%
\providecommand \enquote  [1]{``#1''}%
\providecommand \bibnamefont  [1]{#1}%
\providecommand \bibfnamefont [1]{#1}%
\providecommand \citenamefont [1]{#1}%
\providecommand \href@noop [0]{\@secondoftwo}%
\providecommand \href [0]{\begingroup \@sanitize@url \@href}%
\providecommand \@href[1]{\@@startlink{#1}\@@href}%
\providecommand \@@href[1]{\endgroup#1\@@endlink}%
\providecommand \@sanitize@url [0]{\catcode `\\12\catcode
`\$12\catcode
  `\&12\catcode `\#12\catcode `\^12\catcode `\_12\catcode `\%12\relax}%
\providecommand \@@startlink[1]{}%
\providecommand \@@endlink[0]{}%
\providecommand \url  [0]{\begingroup\@sanitize@url \@url }%
\providecommand \@url [1]{\endgroup\@href {#1}{\urlprefix }}%
\providecommand \urlprefix  [0]{URL }%
\providecommand \Eprint [0]{\href }%
\providecommand \doibase [0]{http://dx.doi.org/}%
\providecommand \selectlanguage [0]{\@gobble}%
\providecommand \bibinfo  [0]{\@secondoftwo}%
\providecommand \bibfield  [0]{\@secondoftwo}%
\providecommand \translation [1]{[#1]}%
\providecommand \BibitemOpen [0]{}%
\providecommand \bibitemStop [0]{}%
\providecommand \bibitemNoStop [0]{.\EOS\space}%
\providecommand \EOS [0]{\spacefactor3000\relax}%
\providecommand \BibitemShut  [1]{\csname bibitem#1\endcsname}%
\let\auto@bib@innerbib\@empty
\bibitem [{\citenamefont {Mak}\ \emph {et~al.}(2010)\citenamefont {Mak},
  \citenamefont {Lee}, \citenamefont {Hone}, \citenamefont {Shan},\ and\
  \citenamefont {Heinz}}]{MakPRL10}%
  \BibitemOpen
  \bibfield  {author} {\bibinfo {author} {\bibfnamefont {K.~F.}\ \bibnamefont
  {Mak}}, \bibinfo {author} {\bibfnamefont {C.}\ \bibnamefont {Lee}},
  \bibinfo {author} {\bibfnamefont {J.}\ \bibnamefont {Hone}}, \bibinfo
  {author} {\bibfnamefont {J.}\ \bibnamefont {Shan}}, \ and\ \bibinfo {author}
  {\bibfnamefont {T.~F.}\ \bibnamefont {Heinz}},\ }\bibfield  {title}
  {\enquote {\bibinfo {title} {Atomically thin {M}o{S}$_{2}$: A new direct-gap
  semiconductor},}\ }\href@noop {} {\bibfield  {journal} {\bibinfo  {journal}
  {Phys. Rev. Lett.}\ }\textbf {\bibinfo {volume} {105}},\ \bibinfo {pages}
  {136805} (\bibinfo {year} {2010})}\BibitemShut {NoStop}%
\bibitem [{\citenamefont {Horng}\ \emph {et~al.}(2019)\citenamefont {Horng},
  \citenamefont {Martin}, \citenamefont {Chou}, \citenamefont {Courtade},
  \citenamefont {Chang}, \citenamefont {Hsu}, \citenamefont {Wentzel},
  \citenamefont {Ruth}, \citenamefont {Lu}, \citenamefont {Cundiff},
  \citenamefont {Wang},\ and\ \citenamefont {Deng}}]{Horng19}%
  \BibitemOpen
  \bibfield  {author} {\bibinfo {author} {\bibfnamefont {J.}~\bibnamefont
  {Horng}}, \bibinfo {author} {\bibfnamefont {E.~W.}\ \bibnamefont {Martin}},
  \bibinfo {author} {\bibfnamefont {Y.-H.}\ \bibnamefont {Chou}}, \bibinfo
  {author} {\bibfnamefont {E.}~\bibnamefont {Courtade}}, \bibinfo {author}
  {\bibfnamefont {T.-C.}\ \bibnamefont {Chang}}, \bibinfo {author}
  {\bibfnamefont {C.-Y.}\ \bibnamefont {Hsu}}, \bibinfo {author} {\bibfnamefont
  {M.-H.}\ \bibnamefont {Wentzel}}, \bibinfo {author} {\bibfnamefont {H.~G.}\
  \bibnamefont {Ruth}}, \bibinfo {author} {\bibfnamefont {T.-C.}\ \bibnamefont
  {Lu}}, \bibinfo {author} {\bibfnamefont {S.~T.}\ \bibnamefont {Cundiff}},
  \bibinfo {author} {\bibfnamefont {F.}~\bibnamefont {Wang}}, \ and\ \bibinfo
  {author} {\bibfnamefont {H.}~\bibnamefont {Deng}},\ }\bibfield  {title}
  {\enquote {\bibinfo {title} {Perfect absorption by an atomically thin
  crystal},}\ }\href@noop {} {\bibfield  {journal} {\bibinfo  {journal}
  {condmat}\ } (\bibinfo {year} {2019})}\BibitemShut {NoStop}%
\bibitem [{\citenamefont {Mak}\ and\ \citenamefont
  {Shan}(2016)}]{MakNatPhot16}%
  \BibitemOpen
  \bibfield  {author} {\bibinfo {author} {\bibfnamefont {K.~F.}\ \bibnamefont
  {Mak}}\ and\ \bibinfo {author} {\bibfnamefont {J.}\ \bibnamefont {Shan}},\
  }\bibfield  {title} {\enquote {\bibinfo {title} {Photonics and
  optoelectronics of 2d semiconductor transition metal dichalcogenides},}\
  }\href@noop {} {\bibfield  {journal} {\bibinfo  {journal} {Nat. Photon.}\ }
  (\bibinfo {year} {2016})}\BibitemShut {NoStop}%
\bibitem [{\citenamefont {Gao}\ \emph {et~al.}(2016)\citenamefont {Gao},
  \citenamefont {Li}, \citenamefont {Tan}, \citenamefont {Chow}, \citenamefont
  {Lu},\ and\ \citenamefont {Koratkar}}]{GaoACSNano16}%
  \BibitemOpen
  \bibfield  {author} {\bibinfo {author} {\bibfnamefont {J.}\ \bibnamefont
  {Gao}}, \bibinfo {author} {\bibfnamefont {B.}\ \bibnamefont {Li}},
  \bibinfo {author} {\bibfnamefont {J.}\ \bibnamefont {Tan}}, \bibinfo
  {author} {\bibfnamefont {P.}\ \bibnamefont {Chow}}, \bibinfo {author}
  {\bibfnamefont {T.-M.}\ \bibnamefont {Lu}}, \ and\ \bibinfo {author}
  {\bibfnamefont {N.}\ \bibnamefont {Koratkar}},\ }\bibfield  {title}
  {\enquote {\bibinfo {title} {Aging of transition metal dichalcogenide
  monolayers},}\ }\href@noop {} {\bibfield  {journal} {\bibinfo  {journal} {ACS
  Nano}\ }\textbf {\bibinfo {volume} {10}},\ \bibinfo {pages} {2628--2635}
  (\bibinfo {year} {2016})}\BibitemShut {NoStop}%
\bibitem [{\citenamefont {Ahn}\ \emph {et~al.}(2016)\citenamefont {Ahn},
  \citenamefont {Kim}, \citenamefont {Nayak}, \citenamefont {Yoon},
  \citenamefont {Lim}, \citenamefont {Shin},\ and\ \citenamefont
  {Shin}}]{AhnACSNano16}%
  \BibitemOpen
  \bibfield  {author} {\bibinfo {author} {\bibfnamefont {S.}\
  \bibnamefont {Ahn}}, \bibinfo {author} {\bibfnamefont {G.}\
  \bibnamefont {Kim}}, \bibinfo {author} {\bibfnamefont {P.~K.}\
  \bibnamefont {Nayak}}, \bibinfo {author} {\bibfnamefont {S.~I.}\
  \bibnamefont {Yoon}}, \bibinfo {author} {\bibfnamefont {H.}\
  \bibnamefont {Lim}}, \bibinfo {author} {\bibfnamefont {H.-J.}\
  \bibnamefont {Shin}}, \ and\ \bibinfo {author} {\bibfnamefont {H.~S.}\
  \bibnamefont {Shin}},\ }\bibfield  {title} {\enquote {\bibinfo {title}
  {Prevention of transition metal dichalcogenide photodegradation by
  encapsulation with h-{BN} layers},}\ }\href@noop {} {\bibfield  {journal}
  {\bibinfo  {journal} {ACS Nano}\ }\textbf {\bibinfo {volume} {10}},\ \bibinfo
  {pages} {8973--8979} (\bibinfo {year} {2016})}\BibitemShut {NoStop}%
\bibitem [{\citenamefont {Moody}\ \emph {et~al.}(2015)\citenamefont {Moody},
  \citenamefont {Kavir~Dass}, \citenamefont {Hao}, \citenamefont {Chen},
  \citenamefont {Li}, \citenamefont {Singh}, \citenamefont {Tran},
  \citenamefont {Clark}, \citenamefont {Xu}, \citenamefont {Bergh\"{a}user},
  \citenamefont {Malic}, \citenamefont {Knorr},\ and\ \citenamefont
  {Li}}]{MoodyNatCom15}%
  \BibitemOpen
  \bibfield  {author} {\bibinfo {author} {\bibfnamefont {G.}\ \bibnamefont
  {Moody}}, \bibinfo {author} {\bibfnamefont {C.}\ \bibnamefont
  {Kavir~Dass}}, \bibinfo {author} {\bibfnamefont {K.}\ \bibnamefont {Hao}},
  \bibinfo {author} {\bibfnamefont {C.-H.}\ \bibnamefont {Chen}},
  \bibinfo {author} {\bibfnamefont {L.-J.}\ \bibnamefont {Li}}, \bibinfo
  {author} {\bibfnamefont {A.}\ \bibnamefont {Singh}}, \bibinfo {author}
  {\bibfnamefont {K.}\ \bibnamefont {Tran}}, \bibinfo {author} {\bibfnamefont
  {G.}\ \bibnamefont {Clark}}, \bibinfo {author} {\bibfnamefont
  {X.}\ \bibnamefont {Xu}}, \bibinfo {author} {\bibfnamefont {G.}\
  \bibnamefont {Bergh\"{a}user}}, \bibinfo {author} {\bibfnamefont {E.}\
  \bibnamefont {Malic}}, \bibinfo {author} {\bibfnamefont {A.}\
  \bibnamefont {Knorr}}, \ and\ \bibinfo {author} {\bibfnamefont {X.}\
  \bibnamefont {Li}},\ }\bibfield  {title} {\enquote {\bibinfo {title}
  {Intrinsic homogeneous linewidth and broadening mechanisms of excitons in
  monolayer transition metal dichalcogenides},}\ }\href@noop {} {\bibfield
  {journal} {\bibinfo  {journal} {Nat. Commun.}\ }\textbf {\bibinfo {volume}
  {7}},\ \bibinfo {pages} {8315} (\bibinfo {year} {2015})}\BibitemShut
  {NoStop}%
\bibitem [{\citenamefont {Jakubczyk}\ \emph {et~al.}(2016)\citenamefont
  {Jakubczyk}, \citenamefont {Delmonte}, \citenamefont {Koperski},
  \citenamefont {Nogajewski}, \citenamefont {Faugeras}, \citenamefont
  {Langbein}, \citenamefont {Potemski},\ and\ \citenamefont
  {Kasprzak}}]{JakubczykNanoLett16}%
  \BibitemOpen
  \bibfield  {author} {\bibinfo {author} {\bibfnamefont {T.}\ \bibnamefont
  {Jakubczyk}}, \bibinfo {author} {\bibfnamefont {V.}\ \bibnamefont
  {Delmonte}}, \bibinfo {author} {\bibfnamefont {M.}\ \bibnamefont
  {Koperski}}, \bibinfo {author} {\bibfnamefont {K.}\ \bibnamefont
  {Nogajewski}}, \bibinfo {author} {\bibfnamefont {C.}\ \bibnamefont
  {Faugeras}}, \bibinfo {author} {\bibfnamefont {W.}\ \bibnamefont
  {Langbein}}, \bibinfo {author} {\bibfnamefont {M.}\ \bibnamefont
  {Potemski}}, \ and\ \bibinfo {author} {\bibfnamefont {J.}\ \bibnamefont
  {Kasprzak}},\ }\bibfield  {title} {\enquote {\bibinfo {title} {Radiatively
  limited dephasing and exciton dynamics in {M}o{S}e$_2$ monolayers revealed
  with four-wave mixing microscopy},}\ }\href@noop {} {\bibfield  {journal}
  {\bibinfo  {journal} {Nano Lett.}\ }\textbf {\bibinfo {volume} {16}},\
  \bibinfo {pages} {5333--5339} (\bibinfo {year} {2016})}\BibitemShut {NoStop}%
\bibitem [{\citenamefont {Cadiz}\ \emph {et~al.}(2017)\citenamefont {Cadiz},
  \citenamefont {Courtade}, \citenamefont {Robert}, \citenamefont {Wang},
  \citenamefont {Shen}, \citenamefont {Cai}, \citenamefont {Taniguchi},
  \citenamefont {Watanabe}, \citenamefont {Carrere}, \citenamefont {Lagarde},
  \citenamefont {Manca}, \citenamefont {Amand}, \citenamefont {Renucci},
  \citenamefont {Tongay}, \citenamefont {Marie},\ and\ \citenamefont
  {Urbaszek}}]{CadizPRX17}%
  \BibitemOpen
  \bibfield  {author} {\bibinfo {author} {\bibfnamefont {F.}~\bibnamefont
  {Cadiz}}, \bibinfo {author} {\bibfnamefont {E.}~\bibnamefont {Courtade}},
  \bibinfo {author} {\bibfnamefont {C.}~\bibnamefont {Robert}}, \bibinfo
  {author} {\bibfnamefont {G.}~\bibnamefont {Wang}}, \bibinfo {author}
  {\bibfnamefont {Y.}~\bibnamefont {Shen}}, \bibinfo {author} {\bibfnamefont
  {H.}~\bibnamefont {Cai}}, \bibinfo {author} {\bibfnamefont {T.}~\bibnamefont
  {Taniguchi}}, \bibinfo {author} {\bibfnamefont {K.}~\bibnamefont {Watanabe}},
  \bibinfo {author} {\bibfnamefont {H.}~\bibnamefont {Carrere}}, \bibinfo
  {author} {\bibfnamefont {D.}~\bibnamefont {Lagarde}}, \bibinfo {author}
  {\bibfnamefont {M.}~\bibnamefont {Manca}}, \bibinfo {author} {\bibfnamefont
  {T.}~\bibnamefont {Amand}}, \bibinfo {author} {\bibfnamefont
  {P.}~\bibnamefont {Renucci}}, \bibinfo {author} {\bibfnamefont
  {S.}~\bibnamefont {Tongay}}, \bibinfo {author} {\bibfnamefont
  {X.}~\bibnamefont {Marie}}, \ and\ \bibinfo {author} {\bibfnamefont
  {B.}~\bibnamefont {Urbaszek}},\ }\bibfield  {title} {\enquote {\bibinfo
  {title} {Excitonic linewidth approaching the homogeneous limit in {M}o{S}$_2$
  based van der {W}aals heterostructures: accessing spin-valley dynamics},}\
  }\href@noop {} {\bibfield  {journal} {\bibinfo  {journal} {Phys. Rev. X}\
  }\textbf {\bibinfo {volume} {7}},\ \bibinfo {pages} {021026} (\bibinfo {year}
  {2017})}\BibitemShut {NoStop}%
\bibitem [{\citenamefont {Ajayi}\ \emph {et~al.}(2017)\citenamefont {Ajayi},
  \citenamefont {Ardelean}, \citenamefont {Shepard}, \citenamefont {Wang},
  \citenamefont {Antony}, \citenamefont {Taniguchi}, \citenamefont {Watanabe},
  \citenamefont {Heinz}, \citenamefont {Strauf}, \citenamefont {Zhu},\ and\
  \citenamefont {Hone}}]{Ayayi2DMater17}%
  \BibitemOpen
  \bibfield  {author} {\bibinfo {author} {\bibfnamefont {O.~A.}\
  \bibnamefont {Ajayi}}, \bibinfo {author} {\bibfnamefont {J.~V.}\
  \bibnamefont {Ardelean}}, \bibinfo {author} {\bibfnamefont {G.~D.}\
  \bibnamefont {Shepard}}, \bibinfo {author} {\bibfnamefont {J.}\ \bibnamefont
  {Wang}}, \bibinfo {author} {\bibfnamefont {A.}\ \bibnamefont
  {Antony}}, \bibinfo {author} {\bibfnamefont {T.}\ \bibnamefont
  {Taniguchi}}, \bibinfo {author} {\bibfnamefont {K.}\ \bibnamefont
  {Watanabe}}, \bibinfo {author} {\bibfnamefont {T.~F.}\ \bibnamefont
  {Heinz}}, \bibinfo {author} {\bibfnamefont {S.}\ \bibnamefont {Strauf}},
  \bibinfo {author} {\bibfnamefont {X.-Y.}\ \bibnamefont {Zhu}}, \ and\ \bibinfo
  {author} {\bibfnamefont {J.~C.}\ \bibnamefont {Hone}},\ }\bibfield  {title}
  {\enquote {\bibinfo {title} {Approaching the intrinsic photoluminescence
  linewidth in transition metal dichalcogenide monolayers},}\ }\href@noop {}
  {\bibfield  {journal} {\bibinfo  {journal} {2D Mater.}\ }\textbf {\bibinfo
  {volume} {4}},\ \bibinfo {pages} {031011} (\bibinfo {year}
  {2017})}\BibitemShut {NoStop}%
\bibitem [{\citenamefont {Manca}\ \emph {et~al.}(2017)\citenamefont {Manca},
  \citenamefont {Glazov}, \citenamefont {Robert}, \citenamefont {Cadiz},
  \citenamefont {Taniguchi}, \citenamefont {Watanabe}, \citenamefont
  {Courtade}, \citenamefont {Amand}, \citenamefont {Renucci}, \citenamefont
  {Marie}, \citenamefont {Wang},\ and\ \citenamefont
  {Urbaszek}}]{MancaNatCom17}%
  \BibitemOpen
  \bibfield  {author} {\bibinfo {author} {\bibfnamefont {M.}~\bibnamefont
  {Manca}}, \bibinfo {author} {\bibfnamefont {M.~M.}\ \bibnamefont {Glazov}},
  \bibinfo {author} {\bibfnamefont {C.}~\bibnamefont {Robert}}, \bibinfo
  {author} {\bibfnamefont {F.}~\bibnamefont {Cadiz}}, \bibinfo {author}
  {\bibfnamefont {T.}~\bibnamefont {Taniguchi}}, \bibinfo {author}
  {\bibfnamefont {K.}~\bibnamefont {Watanabe}}, \bibinfo {author}
  {\bibfnamefont {E.}~\bibnamefont {Courtade}}, \bibinfo {author}
  {\bibfnamefont {T.}~\bibnamefont {Amand}}, \bibinfo {author} {\bibfnamefont
  {P.}~\bibnamefont {Renucci}}, \bibinfo {author} {\bibfnamefont
  {X.}~\bibnamefont {Marie}}, \bibinfo {author} {\bibfnamefont
  {G.}~\bibnamefont {Wang}}, \ and\ \bibinfo {author} {\bibfnamefont
  {B.}~\bibnamefont {Urbaszek}},\ }\bibfield  {title} {\enquote {\bibinfo
  {title} {Enabling valley selective exciton scattering in monolayer {WS}e$_2$
  through upconversion},}\ }\href@noop {} {\bibfield  {journal} {\bibinfo
  {journal} {Nat. Commun.}\ }\textbf {\bibinfo {volume} {8}},\ \bibinfo {pages}
  {14927} (\bibinfo {year} {2017})}\BibitemShut {NoStop}%
\bibitem [{\citenamefont {Jakubczyk}\ \emph {et~al.}(2019)\citenamefont
  {Jakubczyk}, \citenamefont {Nayak}, \citenamefont {Scarpelli}, \citenamefont
  {Masia}, \citenamefont {Liu}, \citenamefont {Dubey}, \citenamefont {Bendiab},
  \citenamefont {Marty}, \citenamefont {Taniguchi}, \citenamefont {Watanabe},
  \citenamefont {Coraux}, \citenamefont {Bouchiat}, \citenamefont {Langbein},
  \citenamefont {Renard},\ and\ \citenamefont {Kasprzak}}]{JakubczykACSNano19}%
  \BibitemOpen
  \bibfield  {author} {\bibinfo {author} {\bibfnamefont {T.}\ \bibnamefont
  {Jakubczyk}}, \bibinfo {author} {\bibfnamefont {G.}\ \bibnamefont
  {Nayak}}, \bibinfo {author} {\bibfnamefont {L.}\ \bibnamefont
  {Scarpelli}}, \bibinfo {author} {\bibfnamefont {F.}\ \bibnamefont
  {Masia}}, \bibinfo {author} {\bibfnamefont {W.-L.}\ \bibnamefont {Liu}},
  \bibinfo {author} {\bibfnamefont {S.}\ \bibnamefont {Dubey}}, \bibinfo
  {author} {\bibfnamefont {N.}\ \bibnamefont {Bendiab}}, \bibinfo {author}
  {\bibfnamefont {L.}\ \bibnamefont {Marty}}, \bibinfo {author}
  {\bibfnamefont {T.}\ \bibnamefont {Taniguchi}}, \bibinfo {author}
  {\bibfnamefont {K.}\ \bibnamefont {Watanabe}}, \bibinfo {author}
  {\bibfnamefont {G.} \bibnamefont {Nogues}}, \bibinfo
  {author} {\bibfnamefont {J.} \bibnamefont {Coraux}}, \bibinfo
  {author} {\bibfnamefont {V.}\ \bibnamefont {Bouchiat}}, \bibinfo
  {author} {\bibfnamefont {W.}\ \bibnamefont {Langbein}}, \bibinfo
  {author} {\bibfnamefont {J.}\ \bibnamefont {Renard}}, \ and\ \bibinfo
  {author} {\bibfnamefont {J.}\ \bibnamefont {Kasprzak}},\ }\bibfield
  {title} {\enquote {\bibinfo {title} {Coherence and density dynamics of
  excitons in a single-layer {M}o{S}$_2$ reaching the homogeneous limit},}\
  }\href@noop {} {\bibfield  {journal} {\bibinfo  {journal} {ACS Nano}\
  }\textbf {\bibinfo {volume} {13}},\ \bibinfo {pages} {3500--3511} (\bibinfo
  {year} {2019})}\BibitemShut {NoStop}%
\bibitem [{\citenamefont {Chernikov}\ \emph {et~al.}(2014)\citenamefont
  {Chernikov}, \citenamefont {Berkelbach}, \citenamefont {Hill}, \citenamefont
  {Rigosi}, \citenamefont {Li}, \citenamefont {Aslan}, \citenamefont
  {Reichman}, \citenamefont {Hybertsen},\ and\ \citenamefont
  {Heinz}}]{ChernikovPRL14}%
  \BibitemOpen
  \bibfield  {author} {\bibinfo {author} {\bibfnamefont {A.}\ \bibnamefont
  {Chernikov}}, \bibinfo {author} {\bibfnamefont {T.~C.}\ \bibnamefont
  {Berkelbach}}, \bibinfo {author} {\bibfnamefont {H.~M.}\ \bibnamefont
  {Hill}}, \bibinfo {author} {\bibfnamefont {A.}\ \bibnamefont {Rigosi}},
  \bibinfo {author} {\bibfnamefont {Y.}\ \bibnamefont {Li}}, \bibinfo
  {author} {\bibfnamefont {O.~B.}\ \bibnamefont {Aslan}}, \bibinfo
  {author} {\bibfnamefont {D.~R.}\ \bibnamefont {Reichman}}, \bibinfo
  {author} {\bibfnamefont {M.~S.}\ \bibnamefont {Hybertsen}}, \ and\ \bibinfo
  {author} {\bibfnamefont {T.~F.}\ \bibnamefont {Heinz}},\ }\bibfield
  {title} {\enquote {\bibinfo {title} {Exciton binding energy and nonhydrogenic
  {R}ydberg series in monolayer {WS}$_2$},}\ }\href@noop {} {\bibfield
  {journal} {\bibinfo  {journal} {Phys. Rev. Lett.}\ }\textbf {\bibinfo
  {volume} {113}},\ \bibinfo {pages} {076802} (\bibinfo {year}
  {2014})}\BibitemShut {NoStop}%
\bibitem [{\citenamefont {Molas}\ \emph {et~al.}(2019)\citenamefont {Molas},
  \citenamefont {Slobodeniuk}, \citenamefont {Nogajewski}, \citenamefont
  {Bartos}, \citenamefont {Bala}, \citenamefont {Babi\ifmmode~\acute{n}\else
  \'{n}\fi{}ski}, \citenamefont {Watanabe}, \citenamefont {Taniguchi},
  \citenamefont {Faugeras},\ and\ \citenamefont {Potemski}}]{MolasPRL19}%
  \BibitemOpen
  \bibfield  {author} {\bibinfo {author} {\bibfnamefont {M.~R.}\ \bibnamefont
  {Molas}}, \bibinfo {author} {\bibfnamefont {A.~O.}\ \bibnamefont
  {Slobodeniuk}}, \bibinfo {author} {\bibfnamefont {K.}~\bibnamefont
  {Nogajewski}}, \bibinfo {author} {\bibfnamefont {M.}~\bibnamefont {Bartos}},
  \bibinfo {author} {\bibfnamefont {\L{}.}\ \bibnamefont {Bala}}, \bibinfo
  {author} {\bibfnamefont {A.}~\bibnamefont {Babi\ifmmode~\acute{n}\else
  \'{n}\fi{}ski}}, \bibinfo {author} {\bibfnamefont {K.}~\bibnamefont
  {Watanabe}}, \bibinfo {author} {\bibfnamefont {T.}~\bibnamefont {Taniguchi}},
  \bibinfo {author} {\bibfnamefont {C.}~\bibnamefont {Faugeras}}, \ and\
  \bibinfo {author} {\bibfnamefont {M.}~\bibnamefont {Potemski}},\ }\bibfield
  {title} {\enquote {\bibinfo {title} {Energy spectrum of two-dimensional
  excitons in a nonuniform dielectric medium},}\ }\href@noop {} {\bibfield
  {journal} {\bibinfo  {journal} {Phys. Rev. Lett.}\ }\textbf {\bibinfo
  {volume} {123}},\ \bibinfo {pages} {136801} (\bibinfo {year}
  {2019})}\BibitemShut {NoStop}%
\bibitem [{\citenamefont {Goryca}\ \emph {et~al.}(2019)\citenamefont {Goryca},
  \citenamefont {Li}, \citenamefont {Stier}, \citenamefont {Taniguchi},
  \citenamefont {Watanabe}, \citenamefont {Courtade}, \citenamefont {Shree},
  \citenamefont {Robert}, \citenamefont {Urbaszek}, \citenamefont {Marie},\
  and\ \citenamefont {Crooker}}]{GorycaNatCom19}%
  \BibitemOpen
  \bibfield  {author} {\bibinfo {author} {\bibfnamefont {M.}~\bibnamefont
  {Goryca}}, \bibinfo {author} {\bibfnamefont {J.}~\bibnamefont {Li}}, \bibinfo
  {author} {\bibfnamefont {A.~V.}\ \bibnamefont {Stier}}, \bibinfo {author}
  {\bibfnamefont {T.}~\bibnamefont {Taniguchi}}, \bibinfo {author}
  {\bibfnamefont {K.}~\bibnamefont {Watanabe}}, \bibinfo {author}
  {\bibfnamefont {E.}~\bibnamefont {Courtade}}, \bibinfo {author}
  {\bibfnamefont {S.}~\bibnamefont {Shree}}, \bibinfo {author} {\bibfnamefont
  {C.}~\bibnamefont {Robert}}, \bibinfo {author} {\bibfnamefont
  {B.}~\bibnamefont {Urbaszek}}, \bibinfo {author} {\bibfnamefont
  {X.}~\bibnamefont {Marie}}, \ and\ \bibinfo {author} {\bibfnamefont {S.~A.}\
  \bibnamefont {Crooker}},\ }\bibfield  {title} {\enquote {\bibinfo {title}
  {Revealing exciton masses and dielectric properties of monolayer
  semiconductors with high magnetic fields},}\ }\href@noop {} {\bibfield
  {journal} {\bibinfo  {journal} {Nat. Commun.}\ }\textbf {\bibinfo {volume}
  {10}},\ \bibinfo {pages} {4172} (\bibinfo {year} {2019})}\BibitemShut
  {NoStop}%
\bibitem [{Fig({\natexlab{a}})}]{FigS1}%
  \BibitemOpen
  \href@noop {} {} \bibinfo {note} {see Supplemental
  Material at [URL will be inserted by publisher] for an example of the photon
  echo formation, in particular in Fig.\,S1\,b.}\BibitemShut {Stop}%
\bibitem [{\citenamefont {Langbein}\ and\ \citenamefont
  {Patton}(2006)}]{LangbeinOL06}%
  \BibitemOpen
  \bibfield  {author} {\bibinfo {author} {\bibfnamefont {W.}~\bibnamefont
  {Langbein}}\ and\ \bibinfo {author} {\bibfnamefont {B.}~\bibnamefont
  {Patton}},\ }\bibfield  {title} {\enquote {\bibinfo {title} {Heterodyne
  spectral interferometry for multidimensional nonlinear spectroscopy of
  individual quantum systems},}\ }\href@noop {} {\bibfield  {journal} {\bibinfo
   {journal} {Opt. Lett.}\ }\textbf {\bibinfo {volume} {31}},\ \bibinfo {pages}
  {1151} (\bibinfo {year} {2006})}\BibitemShut {NoStop}%
\bibitem [{\citenamefont {Langbein}(2010)}]{LangbeinRNC10}%
  \BibitemOpen
  \bibfield  {author} {\bibinfo {author} {\bibfnamefont {W.}~\bibnamefont
  {Langbein}},\ }\bibfield  {title} {\enquote {\bibinfo {title} {Coherent
  optical spectroscopy of semiconductor nanostructures},}\ }\href@noop {}
  {\bibfield  {journal} {\bibinfo  {journal} {Rivista del nuovo cimento}\
  }\textbf {\bibinfo {volume} {33}},\ \bibinfo {pages} {255--312} (\bibinfo
  {year} {2010})}\BibitemShut {NoStop}%
\bibitem [{\citenamefont {Fras}\ \emph {et~al.}(2016)\citenamefont {Fras},
  \citenamefont {Mermillod}, \citenamefont {Nogues}, \citenamefont {Hoarau},
  \citenamefont {Schneider}, \citenamefont {Kamp}, \citenamefont {H\"{o}fing},
  \citenamefont {Langbein},\ and\ \citenamefont {Kasprzak}}]{FrasNatPhot16}%
  \BibitemOpen
  \bibfield  {author} {\bibinfo {author} {\bibfnamefont {F.}~\bibnamefont
  {Fras}}, \bibinfo {author} {\bibfnamefont {Q.}~\bibnamefont {Mermillod}},
  \bibinfo {author} {\bibfnamefont {G.}~\bibnamefont {Nogues}}, \bibinfo
  {author} {\bibfnamefont {C.}~\bibnamefont {Hoarau}}, \bibinfo {author}
  {\bibfnamefont {C.}~\bibnamefont {Schneider}}, \bibinfo {author}
  {\bibfnamefont {M.}~\bibnamefont {Kamp}}, \bibinfo {author} {\bibfnamefont
  {S.}~\bibnamefont {H\"{o}fing}}, \bibinfo {author} {\bibfnamefont
  {W.}~\bibnamefont {Langbein}}, \ and\ \bibinfo {author} {\bibfnamefont
  {J.}~\bibnamefont {Kasprzak}},\ }\bibfield  {title} {\enquote {\bibinfo
  {title} {Multi-wave coherent control of a solid state single emitter},}\
  }\href@noop {} {\bibfield  {journal} {\bibinfo  {journal} {Nat. Phot.}\
  }\textbf {\bibinfo {volume} {10}},\ \bibinfo {pages} {155} (\bibinfo {year}
  {2016})}\BibitemShut {NoStop}%
\bibitem [{\citenamefont {Jakubczyk}\ \emph {et~al.}(2018)\citenamefont
  {Jakubczyk}, \citenamefont {Nogajewski}, \citenamefont {Molas}, \citenamefont
  {Bartos}, \citenamefont {Langbein}, \citenamefont {Potemski},\ and\
  \citenamefont {Kasprzak}}]{Jakubczyk2DMat17}%
  \BibitemOpen
  \bibfield  {author} {\bibinfo {author} {\bibfnamefont {T.}~\bibnamefont
  {Jakubczyk}}, \bibinfo {author} {\bibfnamefont {K.}~\bibnamefont
  {Nogajewski}}, \bibinfo {author} {\bibfnamefont {M.~R.}\ \bibnamefont
  {Molas}}, \bibinfo {author} {\bibfnamefont {M.}~\bibnamefont {Bartos}},
  \bibinfo {author} {\bibfnamefont {W.}~\bibnamefont {Langbein}}, \bibinfo
  {author} {\bibfnamefont {M.}~\bibnamefont {Potemski}}, \ and\ \bibinfo
  {author} {\bibfnamefont {J.}~\bibnamefont {Kasprzak}},\ }\bibfield  {title}
  {\enquote {\bibinfo {title} {Impact of environment on dynamics of exciton
  complexes in a {WS}$_2$ monolayer},}\ }\href@noop {} {\bibfield  {journal}
  {\bibinfo  {journal} {2D Mater.}\ }\textbf {\bibinfo {volume} {5}},\ \bibinfo
  {pages} {031007} (\bibinfo {year} {2018})}\BibitemShut {NoStop}%
\bibitem [{\citenamefont {Fang}\ \emph {et~al.}(2019)\citenamefont {Fang},
  \citenamefont {Han}, \citenamefont {Robert}, \citenamefont {Semina},
  \citenamefont {Lagarde}, \citenamefont {Courtade}, \citenamefont {Taniguchi},
  \citenamefont {Watanabe}, \citenamefont {Amand}, \citenamefont {Urbaszek},
  \citenamefont {Glazov},\ and\ \citenamefont {Marie}}]{FangPRL19}%
  \BibitemOpen
  \bibfield  {author} {\bibinfo {author} {\bibfnamefont {H.~H.}\ \bibnamefont
  {Fang}}, \bibinfo {author} {\bibfnamefont {B.}~\bibnamefont {Han}}, \bibinfo
  {author} {\bibfnamefont {C.}~\bibnamefont {Robert}}, \bibinfo {author}
  {\bibfnamefont {M.~A.}\ \bibnamefont {Semina}}, \bibinfo {author}
  {\bibfnamefont {D.}~\bibnamefont {Lagarde}}, \bibinfo {author} {\bibfnamefont
  {E.}~\bibnamefont {Courtade}}, \bibinfo {author} {\bibfnamefont
  {T.}~\bibnamefont {Taniguchi}}, \bibinfo {author} {\bibfnamefont
  {K.}~\bibnamefont {Watanabe}}, \bibinfo {author} {\bibfnamefont
  {T.}~\bibnamefont {Amand}}, \bibinfo {author} {\bibfnamefont
  {B.}~\bibnamefont {Urbaszek}}, \bibinfo {author} {\bibfnamefont {M.~M.}\
  \bibnamefont {Glazov}}, \ and\ \bibinfo {author} {\bibfnamefont
  {X.}~\bibnamefont {Marie}},\ }\bibfield  {title} {\enquote {\bibinfo {title}
  {Control of the exciton radiative lifetime in van der waals
  heterostructures},}\ }\href@noop {} {\bibfield  {journal} {\bibinfo
  {journal} {Phys. Rev. Lett.}\ }\textbf {\bibinfo {volume} {123}},\ \bibinfo
  {pages} {067401} (\bibinfo {year} {2019})}\BibitemShut {NoStop}%
\bibitem [{\citenamefont {Wegener}\ \emph {et~al.}(1990)\citenamefont
  {Wegener}, \citenamefont {Chemla}, \citenamefont {Schmitt-Rink},\ and\
  \citenamefont {Sch{\"a}fer}}]{WegenerPRA90}%
  \BibitemOpen
  \bibfield  {author} {\bibinfo {author} {\bibfnamefont {M.}~\bibnamefont
  {Wegener}}, \bibinfo {author} {\bibfnamefont {D.~S.}\ \bibnamefont {Chemla}},
  \bibinfo {author} {\bibfnamefont {S.}~\bibnamefont {Schmitt-Rink}}, \ and\
  \bibinfo {author} {\bibfnamefont {W.}~\bibnamefont {Sch{\"a}fer}},\
  }\bibfield  {title} {\enquote {\bibinfo {title} {Line shape of time-resolved
  four-wave mixing},}\ }\href@noop {} {\bibfield  {journal} {\bibinfo
  {journal} {Phys. Rev. A}\ }\textbf {\bibinfo {volume} {42}},\ \bibinfo
  {pages} {5675--83} (\bibinfo {year} {1990})}\BibitemShut {NoStop}%
\bibitem [{\citenamefont {Molina-S\'anchez}\ and\ \citenamefont
  {Wirtz}(2011)}]{MolinaPRB11}%
  \BibitemOpen
  \bibfield  {author} {\bibinfo {author} {\bibfnamefont {A.}~\bibnamefont
  {Molina-S\'anchez}}\ and\ \bibinfo {author} {\bibfnamefont {L.}~\bibnamefont
  {Wirtz}},\ }\bibfield  {title} {\enquote {\bibinfo {title} {Phonons in
  single-layer and few-layer {M}o{S}$_2$ and {WS}$_2$},}\ }\href@noop {}
  {\bibfield  {journal} {\bibinfo  {journal} {Phys. Rev. B}\ }\textbf {\bibinfo
  {volume} {84}},\ \bibinfo {pages} {155413} (\bibinfo {year}
  {2011})}\BibitemShut {NoStop}%
\bibitem [{\citenamefont {Wang}\ \emph {et~al.}(1993)\citenamefont {Wang},
  \citenamefont {Ferrio}, \citenamefont {Steel}, \citenamefont {Hu},
  \citenamefont {Binder},\ and\ \citenamefont {Koch}}]{WangPRL93}%
  \BibitemOpen
  \bibfield  {author} {\bibinfo {author} {\bibfnamefont {H.}\ \bibnamefont
  {Wang}}, \bibinfo {author} {\bibfnamefont {K.}\ \bibnamefont {Ferrio}},
  \bibinfo {author} {\bibfnamefont {D.~G.}\ \bibnamefont {Steel}}, \bibinfo
  {author} {\bibfnamefont {Y.~Z.}\ \bibnamefont {Hu}}, \bibinfo {author}
  {\bibfnamefont {R.}~\bibnamefont {Binder}}, \ and\ \bibinfo {author}
  {\bibfnamefont {S.~W.}\ \bibnamefont {Koch}},\ }\bibfield  {title} {\enquote
  {\bibinfo {title} {Transient nonlinear optical response from excitation
  induced dephasing in {G}a{A}s},}\ }\href@noop {} {\bibfield  {journal}
  {\bibinfo  {journal} {Phys. Rev. Lett.}\ }\textbf {\bibinfo {volume} {71}},\
  \bibinfo {pages} {1261--1264} (\bibinfo {year} {1993})}\BibitemShut {NoStop}%
\bibitem [{\citenamefont {Honold}\ \emph {et~al.}(1989)\citenamefont {Honold},
  \citenamefont {Schultheis}, \citenamefont {Kuhl},\ and\ \citenamefont
  {Tu}}]{HonoldPRB89}%
  \BibitemOpen
  \bibfield  {author} {\bibinfo {author} {\bibfnamefont {A.}~\bibnamefont
  {Honold}}, \bibinfo {author} {\bibfnamefont {L.}~\bibnamefont {Schultheis}},
  \bibinfo {author} {\bibfnamefont {J.}~\bibnamefont {Kuhl}}, \ and\ \bibinfo
  {author} {\bibfnamefont {C.~W.}\ \bibnamefont {Tu}},\ }\bibfield  {title}
  {\enquote {\bibinfo {title} {Collision broadening of two-dimensional excitons
  in a {G}a{A}s single quantum well},}\ }\href@noop {} {\bibfield  {journal}
  {\bibinfo  {journal} {Phys. Rev. B}\ }\textbf {\bibinfo {volume} {9}},\
  \bibinfo {pages} {6442--6445} (\bibinfo {year} {1989})}\BibitemShut {NoStop}%
\bibitem [{Fig({\natexlab{b}})}]{FigS2}%
  \BibitemOpen
  \href@noop {} {} \bibinfo {note} {see Supplemental
  Material at [URL will be inserted by publisher] for corresponding dephasing
  curves measured at the 2S exciton at different temperatures, namely
  Fig.\,S2.}\BibitemShut {Stop}%
\bibitem [{\citenamefont {Brem}\ \emph {et~al.}(2019)\citenamefont {Brem},
  \citenamefont {Zipfel}, \citenamefont {Selig}, \citenamefont {Raja},
  \citenamefont {Waldecker}, \citenamefont {Ziegler}, \citenamefont
  {Taniguchi}, \citenamefont {Watanabe}, \citenamefont {Chernikov},\ and\
  \citenamefont {Malic}}]{BremNanoscale19}%
  \BibitemOpen
  \bibfield  {author} {\bibinfo {author} {\bibfnamefont {S.}\ \bibnamefont
  {Brem}}, \bibinfo {author} {\bibfnamefont {J.}\ \bibnamefont {Zipfel}},
  \bibinfo {author} {\bibfnamefont {M.}\ \bibnamefont {Selig}}, \bibinfo
  {author} {\bibfnamefont {A.}\ \bibnamefont {Raja}}, \bibinfo {author}
  {\bibfnamefont {L.}\ \bibnamefont {Waldecker}}, \bibinfo {author}
  {\bibfnamefont {J.~D.}\ \bibnamefont {Ziegler}}, \bibinfo {author}
  {\bibfnamefont {T.}\ \bibnamefont {Taniguchi}}, \bibinfo {author}
  {\bibfnamefont {K.}\ \bibnamefont {Watanabe}}, \bibinfo {author}
  {\bibfnamefont {A.}\ \bibnamefont {Chernikov}}, \ and\ \bibinfo {author}
  {\bibfnamefont {E.}\ \bibnamefont {Malic}},\ }\bibfield  {title} {\enquote
  {\bibinfo {title} {Intrinsic lifetime of higher excitonic states in tungsten
  diselenide monolayers},}\ }\href@noop {} {\bibfield  {journal} {\bibinfo
  {journal} {Nanoscale}\ }\textbf {\bibinfo {volume} {11}},\ \bibinfo {pages}
  {12381--12387} (\bibinfo {year} {2019})}\BibitemShut {NoStop}%
\bibitem [{\citenamefont {Scarpelli}\ \emph {et~al.}(2017)\citenamefont
  {Scarpelli}, \citenamefont {Masia}, \citenamefont {Alexeev}, \citenamefont
  {Withers}, \citenamefont {Tartakovskii}, \citenamefont {Novoselov},\ and\
  \citenamefont {Langbein}}]{ScarpelliPRB17}%
  \BibitemOpen
  \bibfield  {author} {\bibinfo {author} {\bibfnamefont {L.}~\bibnamefont
  {Scarpelli}}, \bibinfo {author} {\bibfnamefont {F.}~\bibnamefont {Masia}},
  \bibinfo {author} {\bibfnamefont {E.~M.}\ \bibnamefont {Alexeev}}, \bibinfo
  {author} {\bibfnamefont {F.}~\bibnamefont {Withers}}, \bibinfo {author}
  {\bibfnamefont {A.~I.}\ \bibnamefont {Tartakovskii}}, \bibinfo {author}
  {\bibfnamefont {K.~S.}\ \bibnamefont {Novoselov}}, \ and\ \bibinfo {author}
  {\bibfnamefont {W.}~\bibnamefont {Langbein}},\ }\bibfield  {title} {\enquote
  {\bibinfo {title} {Resonantly excited exciton dynamics in two-dimensional
  ${\mathrm{mose}}_{2}$ monolayers},}\ }\href@noop {} {\bibfield  {journal}
  {\bibinfo  {journal} {Phys. Rev. B}\ }\textbf {\bibinfo {volume} {96}},\
  \bibinfo {pages} {045407} (\bibinfo {year} {2017})}\BibitemShut {NoStop}%
\bibitem [{\citenamefont {Becker}\ \emph {et~al.}(2018)\citenamefont {Becker},
  \citenamefont {Scarpelli}, \citenamefont {Nedelcu}, \citenamefont {Raino},
  \citenamefont {Masia}, \citenamefont {Borri}, \citenamefont {Stoferle},
  \citenamefont {Kovalenko}, \citenamefont {Langbein},\ and\ \citenamefont
  {Mahrt}}]{BeckerNanoLett18}%
  \BibitemOpen
  \bibfield  {author} {\bibinfo {author} {\bibfnamefont {M.~A.}\
  \bibnamefont {Becker}}, \bibinfo {author} {\bibfnamefont {L.}\
  \bibnamefont {Scarpelli}}, \bibinfo {author} {\bibfnamefont {G.}\
  \bibnamefont {Nedelcu}}, \bibinfo {author} {\bibfnamefont {G.}\
  \bibnamefont {Raino}}, \bibinfo {author} {\bibfnamefont {F.}\
  \bibnamefont {Masia}}, \bibinfo {author} {\bibfnamefont {P.}\ \bibnamefont
  {Borri}}, \bibinfo {author} {\bibfnamefont {T.}\ \bibnamefont {Stoferle}},
  \bibinfo {author} {\bibfnamefont {M.~V.}\ \bibnamefont {Kovalenko}},
  \bibinfo {author} {\bibfnamefont {W.}\ \bibnamefont {Langbein}}, \ and\
  \bibinfo {author} {\bibfnamefont {R.~F.}\ \bibnamefont {Mahrt}},\
  }\bibfield  {title} {\enquote {\bibinfo {title} {Long exciton dephasing time
  and coherent phonon coupling in {C}s{P}b{B}r$_2${C}l perovskite
  nanocrystals},}\ }\href@noop {} {\bibfield  {journal} {\bibinfo  {journal}
  {Nano Lett.}\ }\textbf {\bibinfo {volume} {18}},\ \bibinfo {pages}
  {7546--7551} (\bibinfo {year} {2018})}\BibitemShut {NoStop}%
\bibitem [{\citenamefont {Naeem}\ \emph {et~al.}(2015)\citenamefont {Naeem},
  \citenamefont {Masia}, \citenamefont {Christodoulou}, \citenamefont
  {Moreels}, \citenamefont {Borri},\ and\ \citenamefont
  {Langbein}}]{NaeemPRB15}%
  \BibitemOpen
  \bibfield  {author} {\bibinfo {author} {\bibfnamefont {A.}\ \bibnamefont
  {Naeem}}, \bibinfo {author} {\bibfnamefont {F.}\ \bibnamefont
  {Masia}}, \bibinfo {author} {\bibfnamefont {S.}\ \bibnamefont
  {Christodoulou}}, \bibinfo {author} {\bibfnamefont {I.}\ \bibnamefont
  {Moreels}}, \bibinfo {author} {\bibfnamefont {P.}\ \bibnamefont {Borri}},
  \ and\ \bibinfo {author} {\bibfnamefont {W.}\ \bibnamefont
  {Langbein}},\ }\bibfield  {title} {\enquote {\bibinfo {title} {Giant exciton
  oscillator strength and radiatively limited dephasing in two-dimensional
  platelets},}\ }\href@noop {} {\bibfield  {journal} {\bibinfo  {journal}
  {Phys. Rev. B}\ }\textbf {\bibinfo {volume} {91}},\ \bibinfo {pages} {121302}
  (\bibinfo {year} {2015})}\BibitemShut {NoStop}%
\bibitem [{\citenamefont {Raja}\ \emph {et~al.}(2019)\citenamefont {Raja},
  \citenamefont {Waldecker}, \citenamefont {Zipfel}, \citenamefont {Cho},
  \citenamefont {Brem}, \citenamefont {Ziegler}, \citenamefont {Kulig},
  \citenamefont {Taniguchi}, \citenamefont {Watanabe}, \citenamefont {Malic},
  \citenamefont {Heinz}, \citenamefont {Berkelbach},\ and\ \citenamefont
  {Chernikov}}]{RajaNatNano19}%
  \BibitemOpen
  \bibfield  {author} {\bibinfo {author} {\bibfnamefont {A.}\ \bibnamefont
  {Raja}}, \bibinfo {author} {\bibfnamefont {L.}\ \bibnamefont {Waldecker}},
  \bibinfo {author} {\bibfnamefont {J.}\ \bibnamefont {Zipfel}}, \bibinfo
  {author} {\bibfnamefont {Y.}\ \bibnamefont {Cho}}, \bibinfo {author}
  {\bibfnamefont {S.}\ \bibnamefont {Brem}}, \bibinfo {author}
  {\bibfnamefont {J.~D.}\ \bibnamefont {Ziegler}}, \bibinfo {author}
  {\bibfnamefont {M.}\ \bibnamefont {Kulig}}, \bibinfo {author}
  {\bibfnamefont {T.}\ \bibnamefont {Taniguchi}}, \bibinfo {author}
  {\bibfnamefont {K.}\ \bibnamefont {Watanabe}}, \bibinfo {author}
  {\bibfnamefont {E.}\ \bibnamefont {Malic}}, \bibinfo {author}
  {\bibfnamefont {T.~F.}\ \bibnamefont {Heinz}}, \bibinfo {author}
  {\bibfnamefont {T.~C.}\ \bibnamefont {Berkelbach}}, \ and\ \bibinfo
  {author} {\bibfnamefont {A.}\ \bibnamefont {Chernikov}},\ }\bibfield
  {title} {\enquote {\bibinfo {title} {Dielectric disorder in two-dimensional
  materials},}\ }\href@noop {} {\bibfield  {journal} {\bibinfo  {journal} {Nat.
  Nanotechnol.}\ }\textbf {\bibinfo {volume} {14}},\ \bibinfo {pages}
  {832--837} (\bibinfo {year} {2019}).}\BibitemShut {Stop}%

\end{thebibliography}
\end{document}